\documentclass[journal]{IEEEtran}
\usepackage[noadjust]{cite}
\usepackage{mathtools}
\usepackage[dvipsnames]{xcolor}
\usepackage{mathrsfs}
\usepackage{amssymb}
\usepackage{algorithm,algorithmicx,algpseudocode}

\usepackage{caption}
\usepackage{subcaption}
\usepackage{tikz}
\usepackage{tikzscale}
\usepackage{pgfplots}
\pgfplotsset{compat=1.8}
\usepackage{xr}
\DeclareMathOperator{\diag}{diag}
\DeclareMathOperator{\prox}{prox}

\newcommand{\F}{\mathcal F}
\newcommand{\samp}{\mathcal S}
\newcommand{\R}{\mathbf R}

\newcommand{\C}{\mathbf C}
\newcommand{\eps}{\varepsilon}
\renewcommand{\geq}{\geqslant}
\renewcommand{\leq}{\leqslant}
\newcommand{\E}[3]{E_{#1}({#2}; {#3})}
\RequirePackage[T1]{fontenc}

\begin{document}

\title{Learning the Sampling Pattern for MRI}

\author{Ferdia~Sherry, Martin~Benning, Juan~Carlos~De~los~Reyes, Martin~J.~Graves, Georg~Maierhofer, Guy~Williams, Carola-Bibiane~Sch\"onlieb and Matthias~J.~Ehrhardt
\thanks{The work of F.~Sherry was supported by the Cantab Capital Institute for the Mathematics of Information. The work of M.~Benning was supported by the Leverhulme Trust Early Career Fellowship 'Learning from mistakes: a supervised feedback-loop for imaging applications'. The work of G.~Maierhofer was supported by the LMS grant URB 16-40. The work of C.-B.~Sch\"onlieb was supported by the Leverhulme Trust project on Breaking the non-convexity barrier, the Philip Leverhulme Prize, the EPSRC grant Nr. EP/M00483X/1, the EPSRC Centre Nr. EP/N014588/1, the European Union Horizon
2020 research and innovation programmes under the Marie Sk\l{}odowska-Curie grant agreement No 777826 NoMADS and
No 691070 CHiPS, the Alan Turing Institute and the Cantab Capital Institute for the Mathematics of Information.    \textit{(Corresponding author: Ferdia Sherry)}}
  \thanks{F.~Sherry, G.~Maierhofer and C.-B.~Sch\"onlieb are with DAMTP, University of Cambridge, Cambridge~CB3~0WA, U.K. (e-mail: fs436@cam.ac.uk).}
  \thanks{M.~Benning is with the School of Mathematical Sciences, QMUL, London~E1~4NS, U.K.}
  \thanks{J.~C.~De los Reyes is with the Research Center on Mathematical Modelling, Escuela Polit\'ecnica Nacional, 170525~Quito, Ecuador.}
  \thanks{M.~J.~Graves is with the Department of Radiology, University of Cambridge, Cambridge~CB2~0QQ, U.K.}
  \thanks{G.~Williams is with the Department of Clinical Neurosciences, University of Cambridge, Cambridge~CB2~0QQ, U.K.}
  \thanks{M.~J.~Ehrhardt is with the IMI, University of Bath, Bath~BA2~7JU, U.K.}
  
}

\maketitle

\begin{abstract}
  The discovery of the theory of compressed sensing brought the realisation that many inverse problems can be solved even when measurements are "incomplete". This is particularly interesting in magnetic resonance imaging (MRI), where long acquisition times can limit its use. In this work, we consider the problem of learning a sparse sampling pattern that can be used to optimally balance acquisition time versus quality of the reconstructed image. We use a supervised learning approach, making the assumption that our training data is representative enough of new data acquisitions. We demonstrate that this is indeed the case, even if the training data consists of just 7 training pairs of measurements and ground-truth images; with a training set of brain images of size 192 by 192, for instance, one of the learned patterns samples only 35\% of k-space, however results in reconstructions with mean SSIM 0.914 on a test set of similar images. The proposed framework is general enough to learn arbitrary sampling patterns, including common patterns such as Cartesian, spiral and radial sampling.
\end{abstract}


\IEEEpeerreviewmaketitle
\section{Introduction}

\IEEEPARstart{T}{he} field of compressed sensing is founded on the realisation that in inverse problems it is often possible to recover signals from incomplete measurements. To do so, the inherent structure of signals and images is exploited. Finding a sparse representation for the unknown signal reduces the number of unknowns and consequently the number of measurements required for reconstruction. This is of great interest in many applications, where external reasons (such as cost or time constraints) typically imply that one should take as few measurements as are required to obtain an adequate reconstruction. A specific example of such an application is magnetic resonance imaging (MRI). In MRI, measurements are modelled as samples of the Fourier transform (points in so-called k-space) of the signal that is to be recovered and taking measurements is a time-intensive procedure. Keeping acquisition times short is important to ensure patient comfort and to mitigate motion artefacts, and it increases patient throughput, thus making MRI effectively cheaper. Hence, MRI is a natural candidate for the application of compressed sensing methodology. While the first theoretical results of compressed sensing (as in \cite{Candes2006}, in which exact recovery results are proven for uniform random sampling strategies) do not apply well to MRI, three underlying principles were identified that enable the success of compressed sensing \cite{Lustig2007},~\cite{Sodickson2015}: 1) sparsity or compressibility of the signal to be recovered (in some sparsifying transform, such as a wavelet transform), 2) incoherent measurements (with respect to the aforementioned sparsifying transform) and 3) a nonlinear reconstruction algorithm that takes advantage of the sparsity structure in the true signal. The nonlinear reconstruction algorithm often takes the form of a variational regularisation problem:
\begin{equation}
  \min_{u} \frac{1}{2}\|\samp\F u - y\|^2 + \alpha R(u),
  \label{prob:general_variational_prob}
\end{equation}
with $\samp$ the subsampling operator, $\F$ the Fourier transform, $y$ the subsampled measurements, $R$ a regularisation functional that encourages the reconstruction to have a sparsity structure and $\alpha$ the regularisation parameter that controls the trade-off between the fit to measurements and fit to structure imposed by $R$. Many previous efforts made towards accelerating MRI have focused on improving how these aspects are treated. The reconstruction algorithm can be changed to more accurately reflect the true structure of the signal: the typical convex reconstruction problem can be replaced by a dictionary learning approach \cite{Ravishankar2011a}; in multi-contrast imaging, structural information obtained from one contrast can be used to inform a regularisation functional to use in the other contrasts \cite{Ehrhardt2016}; and in dynamic MRI additional low rank structure can be exploited to improve reconstruction quality \cite{Lingala2011, Tremoulheac2014}.
\begin{figure}[!htb]
  \centering
  \includegraphics[scale=1]{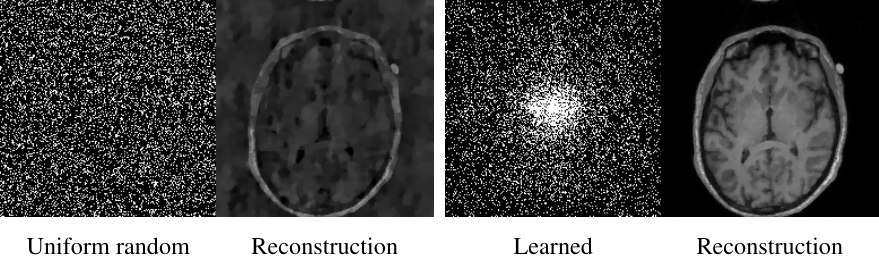}
  \caption{The importance of a good choice of sampling pattern. Left: uniform random pattern (sampling 19\% of k-space) and reconstruction (using total variation type regularisation) on a test image. Right: an equally sparse pattern learned by our algorithm and reconstruction for the same test image.}
  \label{fig:pattern_comparison}
\end{figure}

  It is well known that sampling uniformly at random in k-space (as the original compressed sensing theory suggests \cite{Candes2006}) does not work well in practice; using a variable density sampling pattern greatly improves reconstruction quality \cite{Lustig2007}, see Figure \ref{fig:pattern_comparison}. Note that variable density sampling patterns of scattered points in k-space only allow for accelerated acquisition in 3D, in which case the readout is performed in the orthogonal direction. In the works \cite{Piccini2011, Feng2014, Liu2014, Paquette2015, Usman2009}, subsampling strategies are studied that can be used in practice. On the theoretical side, the compressed sensing assumptions have been refined to derive optimal densities for variable density sampling \cite{Puy2011, Chauffert2013, Chauffert2014}, to prove bounds on reconstruction errors for variable density sampling \cite{Adcock2017, Krahmer2014} and to prove exact recovery results for Cartesian line sampling \cite{Poon2016, Boyer2019}

The sampling pattern can be optimised in a given setting to improve reconstruction quality. There are works on fine-tuning sampling patterns \cite{Ravishankar2011, Seeger2009}, choosing data-adapted sampling patterns without knowledge of the reconstruction method \cite{Knoll2011}, greedy algorithms to pick a suitable pattern for a given reconstruction method \cite{Gozcu2018a, Gozcu2019, Haldar2019}, jointly learning a Cartesian line pattern and neural network reconstruction algorithm \cite{Weiss2019}, and optimal patterns for zero-filling reconstructions can be computed from a training set with little computational effort \cite{Li2016}. We consider the problem of learning an optimal sparse sampling pattern from scratch for a given variational reconstruction method and class of images by solving a bilevel optimisation problem. A similar approach has been used to learn regularisation parameters for variational regularisation models \cite{Delosreyes2017}, among other things.
\subsection{Our Contributions}
\label{sec:contributions}
In this work, we propose a novel bilevel learning approach to learn sparse sampling patterns for MRI. We do this within a supervised learning framework, using training sets of ground truth images with the corresponding measurements.

Our approach can accommodate arbitrary sampling patterns and sampling densities. We demonstrate that the parametrisation of the sampling pattern can be chosen to learn a pattern consisting of a scattered set of points as well as Cartesian lines, but other parametrisations can also be designed that result in radial or spiral sampling, for instance. By using a sparsity promoting penalty on the sampling pattern, we can also vary the sampling rates of our learned patterns.

Besides this, it is also possible to use a wide variety of variational reconstruction algorithms, that is various choices of regularisation $R$ in Problem~\eqref{prob:general_variational_prob}, and we can simultaneously learn the sampling pattern and the optimal regularisation parameter for reconstruction. This forgoes the need to separately tune the parameters of the reconstruction method.

Our optimal sampling patterns confirm empirically the validity of variable density sampling patterns: the optimal patterns tend to sample more densely around the low frequencies and more sparsely at high frequencies. We investigate the dependence of the shape of the sampling density on the sampling rate and the choice of regularisation functional $R$.

By focusing on a particular region within the body, our approach can be used with very small training sets to learn optimal patterns, that nevertheless generalise well to unseen MRI data. We demonstrate this on a set of brain images; indeed, in this setting we find that a training set of just five image, measurement pairs is sufficient.

\section{Model and Methods}
\label{sec:model_intro}
In the bilevel learning framework, the free parameters of a variational regularisation method are learned to optimise a given measure of reconstruction quality. We assume that we are given a variational regularisation method to perform the reconstruction, of a form such as Problem \eqref{prob:general_variational_prob}. Furthermore, we assume that we are given a training set of $N$ pairs of ground truth images $u_i^*$ and fully sampled noisy k-space data $y_i$. With these ingredients we set up a bilevel optimisation problem that can be solved to learn the optimal sampling pattern $\samp$ and regularisation parameter $\alpha$:
  \begin{equation}
    \begin{split}
&      \min_{\samp, \alpha}  \frac{1}{N}\sum\limits_{i=1}^N L_{u_i^*}(\hat u_i(\samp, \alpha)) + P(\samp, \alpha)\\
      &\text{where $\hat u_i(\samp, \alpha)$ solves Problem \eqref{prob:general_variational_prob} with } y=y_i.
    \end{split}
    \label{prob:sampling_bilevel_prob}
  \end{equation}In this problem, we use a continuous parametrisation of the sampling pattern (which is described in detail in Section~\ref{sec:model_upper_level}) so that the learning problem is a continuous optimisation problem. A straightforward generalisation of this parametrisation (which is described in Section~\ref{sec:alt_params} of the Appendix) allows us to impose constraints on the type of pattern that is learned. We will refer to Problem~\eqref{prob:sampling_bilevel_prob} as the upper level problem and will call the variational regularisation problems that make up its constraints the lower level problems. Each $L_{u_i^*}$ is a loss function that quantifies the discrepancy between the reconstruction from subsampled measurements, $\hat u_i$, and the corresponding ground truth $u_i^*$ and $P$ is a penalty on the sampling pattern that encourages its sparsity. Hence, the objective function in Problem~\eqref{prob:sampling_bilevel_prob} is a penalised empirical loss function, the minimiser of which trades off the reconstruction quality against the sparsity of the sampling pattern in an optimal manner. As we show in Section~\ref{sec:diff_ll_sol}, it is possible to differentiate the solution maps $(\samp, \alpha)\mapsto \hat u_i(\samp, \alpha)$ in our setting, so that Problem~\eqref{prob:sampling_bilevel_prob} is amenable to treatment by first order optimisation methods.

In this section, we describe in more detail the various aspects that make up Problem~\eqref{prob:sampling_bilevel_prob} in our setting, starting with the lower level problems, followed by the upper level problem, after which we describe the methods that can be applied to solve the problem.
\subsection{Variational regularisation models}
\label{sec:variational_regularisation}
The lower level problems in Problem~\eqref{prob:sampling_bilevel_prob} are variational regularisation problems. In this section, we specify the class of variational regularisation problems that will be considered. In our application, an image of dimensions $n:= n_1\times n_2$ is modeled as a vector in $\C^n$ by concatenating its columns. The subsampled measurements corresponding to a given image $u$ are modeled as $y=\samp(\F u + \eps)$. Here $\F$ is a Fourier transform operator, $\samp = \diag(s_1,\ldots, s_n), s_i \geq 0$ is the sampling operator, which selects the points in k-space that are included in the measurements (and can be used as a weight on those measurements), and $\eps\in \C^{n}$ is complex Gaussian white noise.

The variational regularisation approach to estimating the true image $u$ from measurements $y$ proceeds by solving an optimisation problem that balances fitting the measurements with fitting prior knowledge that is available about the image. In this work we consider problems that take the form of Problem~\eqref{prob:general_variational_prob} with $R(u) = J(\mathcal Au)$. Here $\mathcal A = (\mathcal A_1,\ldots, \mathcal A_M)$ is a collection of linear operators, $|\mathcal Au|_i = \sqrt{|\mathcal A_1u|_i^2 +\ldots + |\mathcal A_M u|_i^2}$, $\alpha \geq 0$, and $J(v) = \sum_{i=1}^n \rho(|v|_i)$ for some convex $\rho:[0,\infty)\to\R$. Furthermore, we assume that $\rho$ satisfies the following conditions: 1) $\rho$ is increasing, 2) $\rho$ is twice continuously differentiable and 3) $\rho'(u) = \mathcal O(u)$ as $u\to 0$. Finally, a strongly convex penalty $u\mapsto \eps \|u\|^2/2$ is added to the objective function. With these definitions, the lower level energy functional $E_y$, given fully sampled training measurements $y$, takes the following form:
\begin{equation}
    E_y(u;\samp,\alpha) = \frac{1}{2}\|\samp(\F u - y)\|^2 + \alpha J(\mathcal Au) + \frac{\eps}{2}\|u\|^2
    \label{eq:def_ll_energy}
  \end{equation}
  Note that we can approximate a number of common regularisation functionals by choosing $\rho$ to be defined as below for a small $\gamma >0$:
  \[\rho(x) = \begin{cases}-\frac{|x|^3}{3\gamma^2} + \frac{x^2}{\gamma}\quad&\text{if} \quad |x|\leq \gamma\\
      |x| - \frac{\gamma}{3}\quad&\text{if}\quad |x| > \gamma.
    \end{cases}\]
  This choice of $\rho$ can be thought of as a twice continuously differentiable version of the Huber loss function~\cite{Huber1964}. With this $\rho$, we obtain the following types of regularisation:\begin{itemize}
\item if $\mathcal A = \nabla = (\partial_x, \partial_y)$ the regularisation term in Equation~\eqref{eq:def_ll_energy} approximates the isotropic total variation as regularisation term; its use in variational regularisation problems has been studied since~\cite{Rudin1992}, and it is a common choice of regularisation in compressed sensing MRI~\cite{Lustig2007},
\item if $\mathcal A = \mathcal W$ for some sparsifying transform $\mathcal W$, such as a wavelet or shearlet transform, the regularisation term in Equation~\eqref{eq:def_ll_energy} approximates a sparsity penalty on the transform coefficients of the image. These types of regularisation have been successfully applied to compressed sensing MRI in the past~\cite{Guerquin-Kern2009, Pejoski2015}.
\end{itemize}
\subsection{The upper level problem}
\label{sec:model_upper_level}
In the upper level problem, we parametrise the sampling pattern $\samp$ and the lower level regularisation parameter $\alpha$ by a vector $p\in C:= [0,1]^n \times [0,\infty)$: we let $s_i=p_i$ for $i=1,\ldots, n$ and $\alpha(p) = p_{n+1}$. This parametrisation allows us to learn a sampling pattern of scattered points on a grid in k-space, though it is worth noting that the parametrisation can be generalised to constrain the learned pattern. To prevent the notation from becoming overly cumbersome, we do not consider this generalisation here, but refer the reader to Section~\ref{sec:alt_params} in the Appendix for the details.

With this parametrisation, a natural choice of the sparsity penalty $P$ is as follows:
\[P(p) = \beta \sum_{i=1}^{n} p_i + p_i(1-p_i)\]
with $\beta > 0$ a parameter that decides how reconstruction quality is traded off against sparsity of the sampling pattern. Besides encouraging a sparse sampling pattern, this penalty encourages the weights in the sampling pattern $\samp(p)$ to take either the value 0 or 1. For the loss function $L$, we choose $L_{ u'}(u) = \frac{1}{2}\|u - u'\|^2$, but it is straightforward to replace this by any other smooth loss function. For instance, if one is interested in optimising the quality of the recovered edges one could use the smoothed total variation as a loss function: $L_{u'}(u) = \sum_{i=1}^n h_\gamma (|\nabla u' - \nabla u|_i)$, with $h_\gamma$ as defined in Section~\ref{sec:variational_regularisation}.


\subsection{Methods}
As was mentioned in Section~\ref{sec:model_intro}, first order optimisation methods can be used to solve problems like Problem~\eqref{prob:sampling_bilevel_prob}, provided that the solution maps of the lower level problems, $p\mapsto \hat u_i(p)$, can be computed and can be differentiated. In this section we describe the approach taken to computing the solution maps and their derivatives and then describe how these steps are combined to apply first order optimisation methods to Problem~\eqref{prob:sampling_bilevel_prob}.
\subsubsection{Computing the solution maps of the lower level problems}
\label{sec:method_ll}
In this and the next subsection, we will consider the lower level problem for a fixed $y$, so for the sake of notational clarity, we will drop the subscript and write $E=E_y$. The lower level energy functional $E$ is convex in $u$ and takes the saddle-point structure that is used in the primal-dual hybrid gradient algorithm (PDHG) of Chambolle and Pock~\cite{Chambolle2011}. Indeed, we can write
\[E(u;\samp(p), \alpha(p)) = F(\mathcal Ku) + G(u),\]
with $F(v) = F_1(v_1) + F_2(v_2)$, $\mathcal K = (\mathcal K_1, \mathcal K_2)$, where $\mathcal K_1=I$, $\mathcal K_2 = \mathcal A$ and
\begin{align*}
  F_1(v_1) &= \frac{1}{2} \|\samp(p)(\F v_1 - y)\|^2,\\ F_2(v_2) &= \alpha(p) J(v_2),\\ G(u)& =\frac{\eps}{2}\|u\|^2.
\end{align*}
  With this splitting, the parameter choices from Section~\ref{sec:ll_details} in the Appendix and an arbitrary initialisation $u^0$ (we can take it to be the zero-filling reconstruction, or warm start the solver) the following iterative algorithm solves the lower level problem with a linear convergence rate:
  \begin{algorithm}
    \caption{Solving the lower level problem, Problem~\eqref{prob:general_variational_prob}, with PDHG}
    \label{alg:lower_level_solver}
    \begin{algorithmic}
      \Require $u^0, \texttt{maxit}, \texttt{tol}$
      \State $v^0 \leftarrow \mathcal K u^0$
      \State $ \overline u^0\leftarrow u^0$
      \For {$k=0$ to \texttt{maxit}}
      \State $v^{k+1}\leftarrow \prox_{\sigma F^*}(v^k + \mathcal K \overline u^k)$
      \State $u^{k+1} \leftarrow \prox_{\tau G}(u^k - \tau\mathcal K^*v^{k+1})$
      \State $\overline{u}^{k+1} = u^{k+1} + \theta (u^{k+1} - u^k)$
      \If {$\frac{\|u^{k+1} - u^k\|}{\|u^k\|}+\frac{\|v^{k+1} - v^k\|}{\|v^k\|} \leq \texttt{tol}$}
      \State \textbf{break the loop}
      \EndIf
      \EndFor
      \Ensure $u^{k+1}$
    \end{algorithmic}
    \end{algorithm}
    \subsubsection{Differentiating the solution map}
    \label{sec:diff_ll_sol}
    In the previous subsection, we saw that we can compute the solution maps of the lower level problems. To apply first order optimisation methods to Problem~\eqref{prob:sampling_bilevel_prob}, we still need to be able to differentiate these solution maps. To this end, note that the solution map $\hat u$ of $E$ can be defined equivalently by its first order optimality condition:
    \[D_u E(\hat u(p);p) = 0\]
    and that $E$ is twice continuously differentiable in our setting. To ease notation, let us write $\hat u_p:= \hat u(p)$ in this subsection. Since $E$ is strongly convex in $u$, its Hessian is positive definite and hence invertible. As a consequence, the implicit function theorem tells us that the optimality condition can be implicitly differentiated with respect to $p$ and solved to give the derivative of the solution map:
   \[D^2_uE(\hat u_p;p) D_p\hat u_p + D_{u, p}E(\hat u_p;p) = 0, \]
    so that
      \begin{equation}
        \label{eq:grad_loss_function}
        D_p\hat u_p = -[D^2_u E(\hat u_p;p)]^{-1} D_{u, p}E(\hat u_p;p).
      \end{equation}
      In fact, we do not need the full derivative of the solution map in our application, but just the gradient of a scalar function of the solution map, namely $p \mapsto  L_{u^*}(\hat u_p)$ for some ground truth $u^*$. The chain rule and the formula in Equation~\eqref{eq:grad_loss_function} give us a formula for this gradient:
      \begin{equation}
      \begin{split}
        g &=  \nabla_{\hat u_p} L_{u^*}(\hat u_p) D_p\hat u_p\\
        &=-\nabla_{\hat u_p} L_{ u^*}(\hat u_p) [D^2_uE(\hat u_p;p)]^{-1} D_{u, p}E(\hat u_p;p) \\
        &=- D_{p, u} E(\hat u_p;p)[D_u^2E(\hat u_p;p)]^{-1} \nabla_{\hat u_p} L_{ u^*}(\hat u_p)^*.
     \end{split}
     \label{eq:sol_map_grad}
   \end{equation}
   It is worth noting that this expression for the gradient can also be derived using the Lagrangian formulation of Problem~\eqref{prob:sampling_bilevel_prob}, through the adjoint equation, and this is the way in which it is usually derived when an optimal control perspective is taken~\cite{Delosreyes2017}. To implement this formula in practice, we do not compute the Hessian matrix of $E$ and invert it exactly (since the Hessian is very large; it has as many rows and columns as the images we are dealing with have pixels). Instead, we emphasise that the Hessian is symmetric positive definite, so that it is suitable to solve the linear system with an iterative solver such as the conjugate gradient method. For this, we just need to compute the action of the Hessian, for which we can give explicit expressions. These computations have been done in Section~\ref{sec:hessian_details} of the appendix. The expressions derived in the appendix for $D^2_u E$ and $D_{p,u} E$ can be implemented efficiently in practice and are then used in the conjugate gradient method (CG) to compute the desired gradients.
 \subsubsection{Solving the bilevel problem using L-BFGS-B}
Recall that we are interested in solving Problem~\eqref{prob:sampling_bilevel_prob}. By the previous sections, we know that the objective function of this problem is continuously differentiable, and the constraints that we impose on the parameters form a box constraint, so the optimisation problem that we consider is amenable to treatment by the L-BFGS-B algorithm~\cite{Byrd1995a, Zhu1997}. In our description of the computation of the objective function value and gradient of Problem~\eqref{prob:sampling_bilevel_prob}, we will denote the gradient of $p\mapsto L_{u_i^*}(\hat u_i(p))$ by $g_i$. Since the objective function splits as a sum over the training set, it is completely straightforward to parallelise the computations of the solution maps and desired gradients over the training set:
 \begin{algorithm}[H]
   \caption{Computing the objective function value $L$ and gradient $g$ of the bilevel problem, Problem~\eqref{prob:sampling_bilevel_prob}, at $p$}
   \label{alg:compute_obj_grad}
   \begin{algorithmic}
     \Require $p$
     \For {$i=1$ to $N$}
     \State Set measurements for training example $i$: $y\leftarrow y_i$
     \State Set current $\samp$ and $\alpha$: $\samp \leftarrow\samp(p), \alpha\leftarrow\alpha(p)$
     \State Solve Problem~\eqref{prob:general_variational_prob} with Algorithm~\ref{alg:lower_level_solver} to obtain $\hat u_i$
     \State Solve the system in Equation~\eqref{eq:sol_map_grad} with CG to obtain $g_i$
     \EndFor
     \State $L \leftarrow \frac{1}{N}\sum_{i=1}^N L_{u_i^*}(\hat u_i) + P(p)$
     \State $g \leftarrow \frac{1}{N}\sum_{i=1}^N g_i + \nabla P(p)$
     \Ensure $L, g$
   \end{algorithmic}
 \end{algorithm}
The output of algorithm~\ref{alg:compute_obj_grad} can be plugged in to L-BFGS-B to solve Problem~\eqref{prob:sampling_bilevel_prob}.

\section{Experiments}
Our methods have been implemented in Python, using the PyTorch package \cite{Paszke2019} to solve the lower level problems and adjoint equations (Equation \eqref{eq:sol_map_grad}). We implement the lower level solver as a custom PyTorch module with the backpropagation given by solving the adjoint equation, which allows it to be easily used as a component in another machine learning problem and enables us to make use of GPUs to accelerate computations if available. Our code is available at \texttt{https://github.com/fsherry/bilevelmri}. We use the implementation of the L-BFGS-B algorithm that is included in SciPy \cite{2020SciPy-NMeth} and a PyTorch implementation of the discrete wavelet transform~\cite{fergal_cotter_fbcotterpytorch_wavelets_2019} for our experiments involving wavelet regularisation. All experiments were run on a computer with an Intel Xeon Gold 6140 CPU and a NVIDIA Tesla P100 GPU. Since the learning problem is non-convex, care must be taken with the choice of initialisation. In the experiments in this section, we initialise the learning with a full sampling pattern and the corresponding optimal regularisation parameter. This optimal regularisation parameter is learned using our method, keeping the sampling pattern fixed to fully sample k-space; the optimal regularisation parameter is typically found in less than 10 iterations of the L-BFGS-B algorithm. In practice, this initialisation is found to work well.

In this section, we have experiments in which we look at

- varying the sparsity parameter $\beta$ to control the sparsity of the learned pattern,

- learning Cartesian line patterns with our method,

- using different lower level regularisations,

- varying the size of the training set,

- comparing the learned patterns to other sampling patterns,

- learning sampling patterns for high resolution imaging.

Unless otherwise specified, we use a total variation type regularisation in the lower level problems for all experiments. That is, $\rho$ is chosen as the Huber type function defined in Section~\ref{sec:variational_regularisation} and $\mathcal A = \nabla$. We refer the reader to the supporting document for figures that may be of interest, but are not crucial to the understanding of the results.
\subsection{Data}
The brain images are of size $192\times 192$, taken as slices from 7 separate T1-weighted 3D scans. The corresponding noisy measurements are simulated by taking discrete Fourier transforms of these slices and adding complex Gaussian white noise. In all experiments except the one in Section~\ref{sec:vary_training}, we use a training set consisting of 7 slices. We use 70 slices different to those used in training to test the performance of learned patterns. The scans were acquired on a Siemens PrismaFit scanner. For all scans except one, TE = 2.97~ms, TR = 2300~ms and the Inversion Time was 1100~ms. For the other scan, TE = 2.98~ms, TR = 2300~ms and the Inversion Time was 900~ms.

The brain images used in the experiments shown in Figure~\ref{fig:dtv_comparison} are of size $217\times 181$, taken as slices from a simulated T2-weighted 3D scan from the BrainWeb database~\cite{Cocosco1997}. Noisy measurements are simulated from these slices by taking discrete Fourier transforms and adding complex Gaussian white noise. We use a training set consisting of 5 slices and we use 5 slices different to those used in training to test the performance of learned patterns. In these experiments, the corresponding slices from the T1-weighted scan are used to inform the directional vector fields that are used in the directional total variation regularisation~\cite{Ehrhardt2016} in the lower level problems.

The high resolution images are of size $1024\times 1024$, taken as slices from a T1-weighted 3D scan of a test phantom. We use a training set consisting of 5 slices and test the learned pattern on a single slice different to the ones used in training. Again, the noisy measurements are simulated by taking discrete Fourier transforms of these slices and adding complex Gaussian white noise. The scan was acquired on a GE 3T scanner using spoiled gradient recalled acquisition with TE = 12~ms and TR = 37~ms.
\subsection{Varying the sparsity parameter $\beta$}
\label{sec:vary_beta}
\begin{figure}[!htb]\centering
  \includegraphics[scale=1]{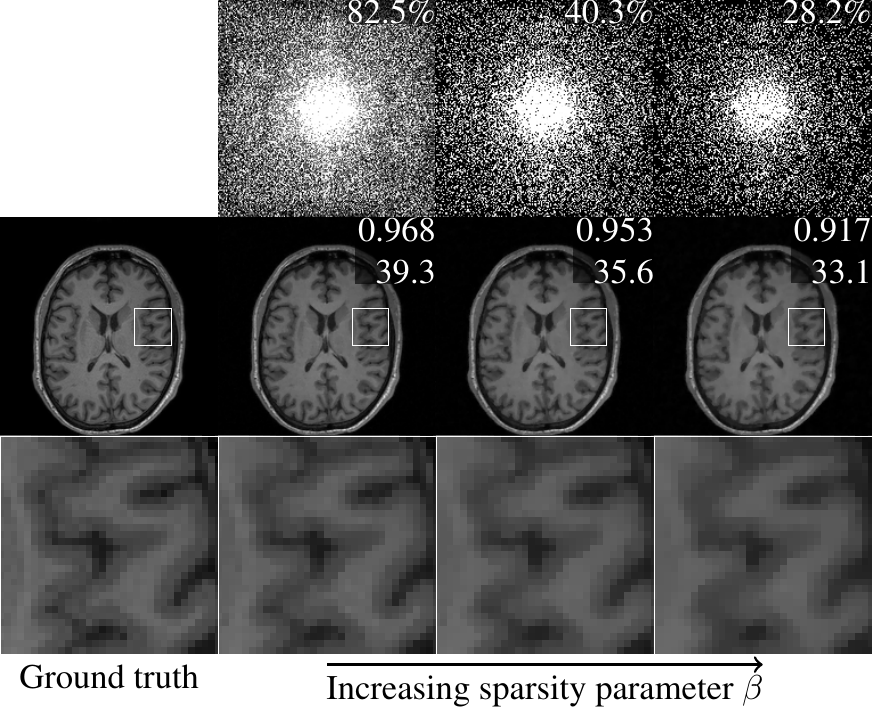}
  \caption{Learned sampling patterns and the corresponding reconstructions on a test image with TV regularisation in the lower level problem. On each of the reconstructions, the top number is the SSIM value and the bottom number is the PSNR. The values of $\beta$ used were (from left to right) $1.58\cdot 10^{-4}, 1.58\cdot 10^{-3}, 1.58\cdot 10^{-2}$.}
  \label{fig:tv_patterns_recons}
  \end{figure}

\begin{figure}[!htb]\centering
    \includegraphics[scale=1]{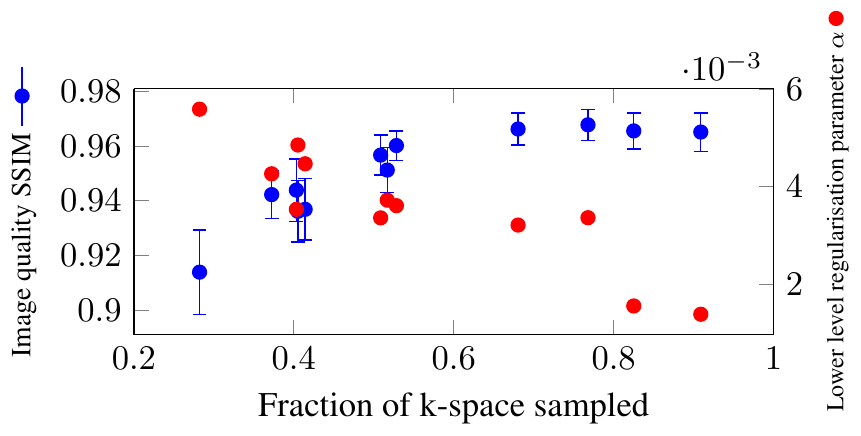}
    \caption{Performance of the learned patterns (measured using the SSIM index) on the test set, and the lower level regularisation parameter $\alpha$ that was learned, against the fraction of k-space that is sampled.}
    \label{fig:tv_ssims_alphas}
  \end{figure}

\begin{figure}[!htb]\centering
    \includegraphics[scale=1]{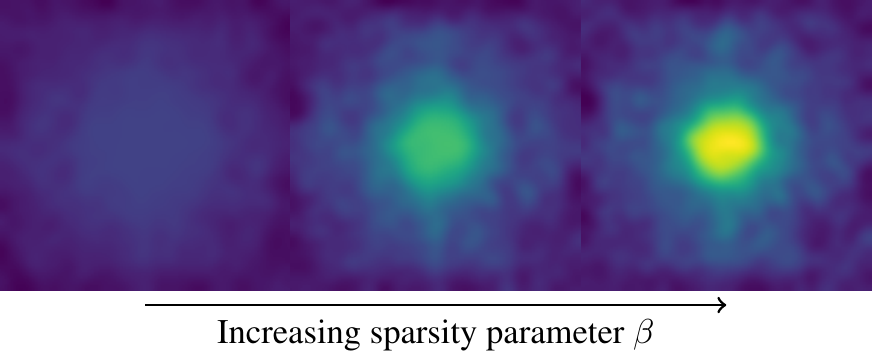}
    \caption{Gaussian kernel density estimates of the sampling distributions for reconstruction with TV regularisation.}
    \label{fig:tv_kde}
    \end{figure}

Learning with a training set of 7 brain images, we consider the effect of varying the sparsity parameter $\beta$. Increasing this parameter tends to make the learned patterns sparser, although we do see a slight deviation from this monotone behaviour for large $\beta$. Figure~\ref{fig:tv_patterns_recons} shows examples of the learned patterns and reconstructions on a test image and in Figure~\ref{fig:tv_ssims_alphas}, we see the performance of the learned patterns, evaluated on the test set of 70 brain images.
We use a Gaussian kernel density estimator to estimate a sampling distribution corresponding to each pattern. That is, we convolve the learned pattern with a Gaussian filter with a small bandwidth and normalise the resulting image to sum to 1. The results of doing this can be seen in Figure~\ref{fig:tv_kde}: we see that the distributions become more peaked strongly around the origin as the patterns become sparser and furthermore, we see that the decay in the learned patterns is anisotropic (as opposed to the isotropic decay of variable density sampling patterns that are not adapted to the data, such as in~\cite{Lustig2007}). 

    \subsection{Cartesian line sampling}
    \label{sec:cartesian_line}
\begin{figure}[!htb]\centering
  \includegraphics[scale=1]{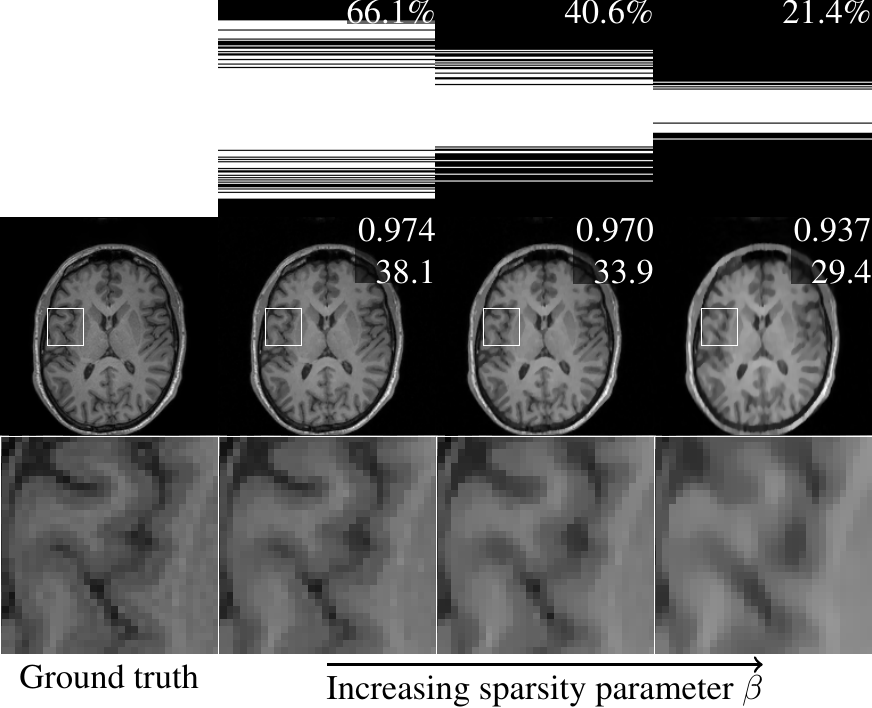}
  \caption{Learned Cartesian line sampling patterns and the corresponding reconstructions on a test image with TV regularisation in the lower level problem. On each of the reconstructions, the top number is the SSIM value and the bottom number is the PSNR. The values of $\beta$ used were (from left to right) $1.58\cdot 10^{-3}, 6.31\cdot 10^{-3}, 1.58\cdot 10^{-2}$.}
  \label{fig:line_patterns_recons}
\end{figure}

As described in Section~\ref{sec:alt_params} of the Appendix, we can restrict the learned pattern to sample along Cartesian lines. Similarly to the case of learning scattered points in k-space, we see in Figure~\ref{fig:line_patterns_recons} that we have some control over the sparsity of the learned pattern using the parameter $\beta$. The sparsity penalty $P$ does not seem to work as well in this situation in encouraging the weights of the pattern to be binary, so we threshold the resulting patterns (that is, we take $p^\text{thresholded}_i = 1$ if $p_i > 0$ and $p^\text{thresholded}_i = 0$ if $p_i=0$) and tune the lower level regularisation parameter on the training set using our method and the thresholded pattern. 
  \subsection{Other lower level regularisations}
  \subsubsection{Wavelet regularisation}
 \label{sec:wavelet}
\begin{figure}[!htb]
    \centering
    \raisebox{-0.5\height}{\includegraphics[scale=1]{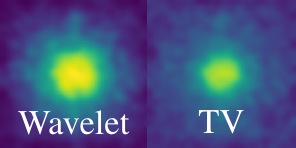}}
    \raisebox{-0.5\height}{\includegraphics[scale=1]{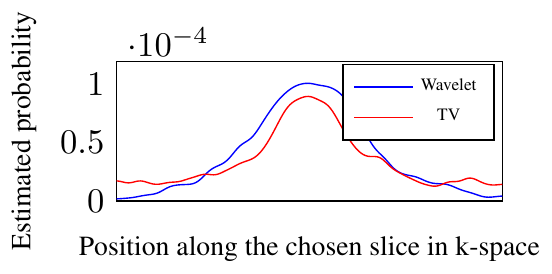}}
        \caption{Gaussian kernel density estimates of the sampling distributions for reconstruction with wavelet and TV regularisation (for approximately the same sparsity in k-space). On the right we plot slices taken along the diagonal of these distributions, showing clearly that the sampling distribution for reconstruction with wavelet regularisation is more strongly peaked around the centre.}
    \label{fig:tv_wavelet_kde}
    \end{figure}

Instead of the TV type regularisation, we use a sparsity penalty on the wavelet coefficients of the image. We accomplish this by choosing $\rho = h_\gamma$ and $\mathcal A=\mathcal W$ for $\mathcal W$ an orthogonal wavelet transform (we use Daubechies 4 wavelets). This results in learned sampling patterns that have slightly different qualititative properties compared to those for the total variation regularisation. Comparing two patterns from the TV and wavelet regularisation with the same sparsity, we find that the pattern for the wavelet regularisation is more strongly peaked around the origin. We can see this in Figure~\ref{fig:tv_wavelet_kde}, where we have estimated the sampling distributions for two learned patterns with TV and wavelet regularisation, both of which sample approximately 27\% of k-space.
    \subsubsection{$H^1$ regularisation}
\begin{figure}[!htb]\centering
    \includegraphics[scale=1]{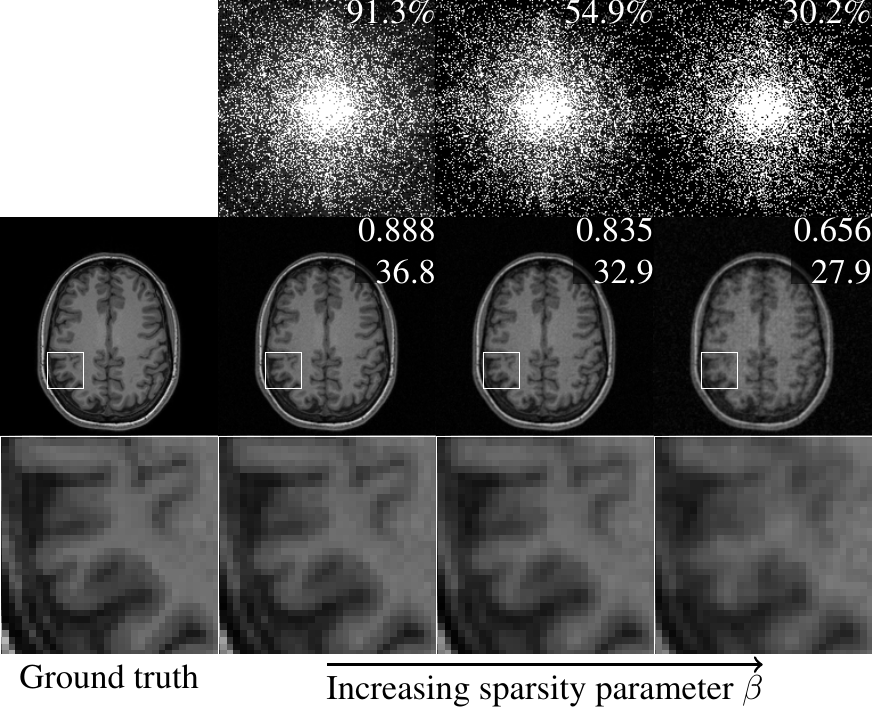}
    \caption{Learned sampling patterns and the corresponding reconstructions on a test image with $H^1$ regularisation in the lower level problem. On each of the reconstructions, the top number is the SSIM value and the bottom number is the PSNR. The values of $\beta$ used were (from left to right) $ 10^{-3}, 2.51\cdot 10^{-3}, 6.31\cdot 10^{-3}$.}
    \label{fig:tikhonov_patterns_recons}
  \end{figure}

    We use the squared $H^1$ seminorm as lower level regularisation, if we take $\rho(x) = x^2/2$ and $\mathcal A= \nabla$ in the lower level problem. With this choice, we find that the learned $\alpha$ equals $0$ and that the learned pattern does not take on just binary values: the weights of the learned pattern are lower at higher frequencies, as can be seen in Figure~\ref{fig:tikhonov_patterns_recons}.

  \subsubsection{No regularisation}
 \label{sec:noreg}
Taking no regularisation in the lower level problem, i.e.\! $\rho =0$ and fixing $\alpha = 0$, we find essentially the same results as when we considered the $H^1$ regularisation: the weights in the learned pattern show a decay away from the origin as in Figure~\ref{fig:tikhonov_patterns_recons}.
\subsubsection{Comparison of the different regularisations}
\begin{figure}[!htb]\centering
  \includegraphics[scale=1]{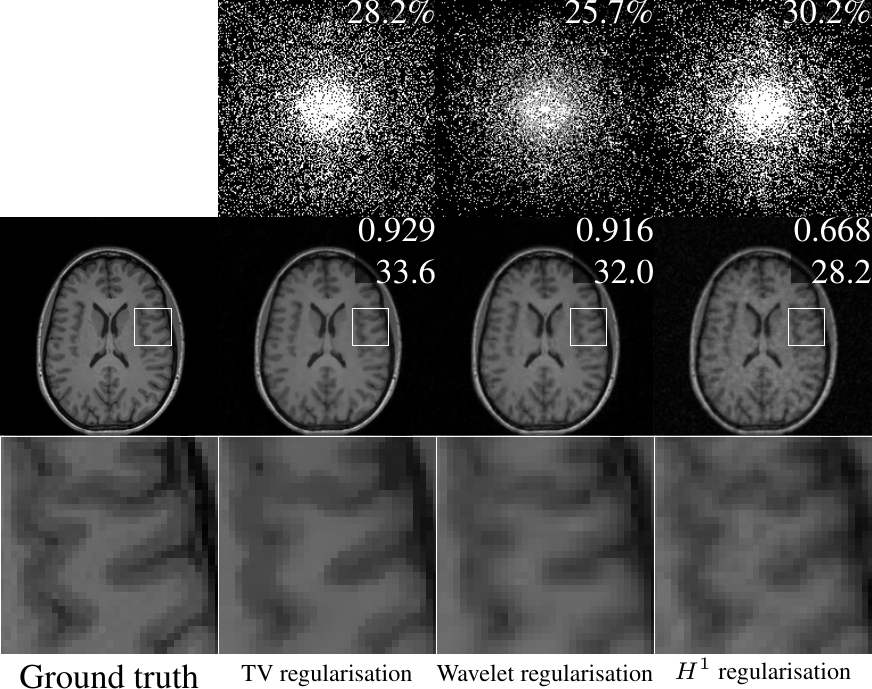}
  \caption{A comparison of learned sampling patterns for the different lower level regularisations that we have considered. On each of the reconstructions, the top number is the SSIM value and the bottom number is the PSNR.}
  \label{fig:compare_regs}
  \end{figure}

We compare the performance of the learned patterns with the different lower level regularisations. In Table~\ref{tab:compare_regs}, we list the performance of three of these patterns on the test set of brain images, each pattern sampling roughly the same proportion of k-space.
\begin{table}[!h]
  \caption{Performance of the learned patterns with different lower level regularisation functionals.}
  \label{tab:compare_regs}
  \centering

    \begin{tabular}{|l|l|l|l|}
      \hline
      &Regularisation & SSIM & PSNR\\
      \hline
      \textbf{Training}&TV (28.2\%) & $0.980\pm0.002$ & $31.6\pm0.5$\\
      &Wavelet (25.7\%) & $0.962\pm 0.003$ & $29.3\pm 0.4$\\
      &$H^1$ (30.2\%) & $0.872\pm 0.004$ & $25.9\pm 0.3$\\
      \hline
       \textbf{Testing}&TV (28.2\%) & $0.915\pm 0.002$ & $33.1\pm 0.7$\\
       &Wavelet (25.7\%)& $0.913 \pm 0.001$ & $31.9 \pm 0.7$\\
       &$H^1$ (30.2\%) & $0.651 \pm 0.005$ & $28.1\pm 0.5$\\
      \hline
    \end{tabular}

  \end{table}  
The TV regularisation is seen to outperform wavelet regularisation, which in turn outperforms $H^1$ regularisation. Figure~\ref{fig:compare_regs} shows the three patterns that we are comparing and the corresponding reconstructions on a test image. We note that this method can easily be extended to other regularisation functions (such as the Total Generalised Variation) that have been used in the context of MRI~\cite{Knoll2011, Benning2014}.

  \subsection{Varying the size of the training set}
  \label{sec:vary_training}
\begin{figure}[!htb]\centering
    \includegraphics[scale=1]{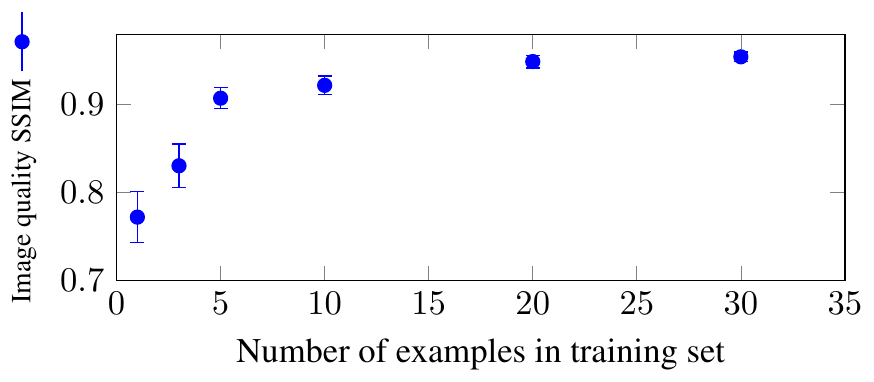}
    \caption{The performance of the learned pattern on the test set as it depends on the size of the training set.}
    \label{fig:plot_ssim_trainingset}
  \end{figure}

To investigate the effect of the size of the training set, we ran our method on different training sets of slices of brain images, of sizes 1, 3, 5, 10, 20, 30 to obtain sampling patterns of roughly the same sparsity. As we see in Figure~\ref{fig:plot_ssim_trainingset}, the learned patterns perform reasonably well (on the training set of 70 slices) from a training set of size 5 and performance flattens out as the size of the training set increases to about 20.
  \subsection{Comparing with other patterns}
  \label{sec:compare_patterns}
In this subsection, we compare the performance of our learned patterns to the performance of sampling patterns chosen using other strategies. Section~\ref{sec:compare_free_patterns} considers the problem of choosing a sampling pattern of scattered 2D points, while Section~\ref{sec:compare_line_patterns} discusses the case where sampling is constrained to Cartesian lines.
    \subsubsection{Free patterns}
    \label{sec:compare_free_patterns}
\begin{figure}[!htb]
        \centering
        \includegraphics{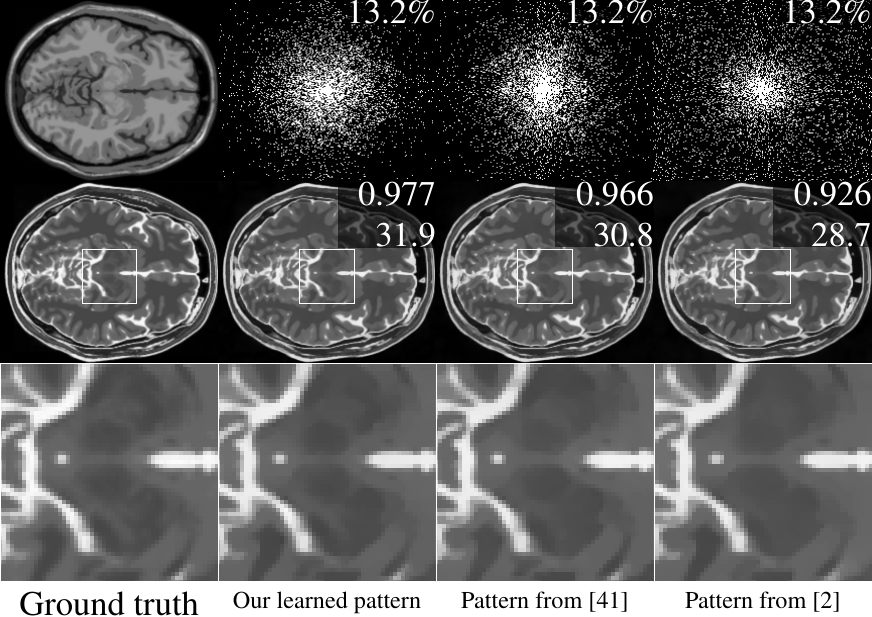}
        \caption{A comparison of our learned pattern to another data-adapted pattern~\cite{Knoll2011a} and an uninformed variable density sampling pattern~\cite{Lustig2007} with dTV regularisation in the lower level problem. The example image shown is a test example, not seen by our learned method or the data-adapted method at training time. On each of the reconstructions, the top number is the SSIM value and the bottom number is the PSNR. The top image in the ground truth column is the T1-weighted slice that is used to generate the reference vector field for the dTV regularisation for this test example.}
        \label{fig:dtv_comparison}
      \end{figure}

    We compare our method for learning sampling patterns to a different data-adapted method for generating sampling patterns~\cite{Knoll2011a} and to uninformed variable density sampling patterns as in~\cite{Lustig2007}. In this comparison, we use directional total variation regularisation~\cite{Ehrhardt2016} in the lower level problem. We use slices from a T1-weighted 3D scan from the BrainWeb database~\cite{Cocosco1997} to generate reference vector fields and use the corresponding slices from the T2-weighted scan as ground truths. The pattern is learned with a training set of 5 slices and checked on a testing set of 5 slices. Neither the data-adapted pattern from~\cite{Knoll2011a} nor the uninformed variable density sampling pattern from~\cite{Lustig2007} fix the lower level regularisation parameter, so we fix these by using our method to learn the optimal regularisation parameter on the training set. The directional total variation is a strong form of regularisation since edge information from one modality is used to regularise the reconstruction of another modality. As a result, we see in Figure~\ref{fig:dtv_comparison} and Table~\ref{tab:compare_dtv} that reconstructions with all of the patterns are relatively good, even at a low sampling rate. Comparing the details we see that both of the data-adapted patterns outperform the uninformed variable density sampling pattern, and that our learned pattern outperforms both other patterns. Since our pattern was learned using knowledge of the lower level regularisation and the pattern from~\cite{Knoll2011a} does not use this information, we conclude that it is possible to adapt to the reconstruction method to improve sampling strategies. The zoomed regions in Figure~\ref{fig:dtv_comparison} show that our method does a better job at resolving the fine structures in the image.

    \begin{table}[!h]
  \caption{A comparison of the performance of our learned pattern to the data-adapted patterns of~\cite{Knoll2011a} and uninformed variable density sampling patterns from~\cite{Lustig2007} with dTV regularisation in the lower level problem. All compared sampling patterns sample 13.2\% of k-space.}
  \label{tab:compare_dtv}
  \centering

    \begin{tabular}{|l|l|l|l|}
      \hline
      &Pattern type& SSIM & PSNR\\
      \hline
      \textbf{Training}&Our method & $0.977\pm 0.002$ & $32.5\pm 0.2$\\
      &Data-adapted~\cite{Knoll2011a}  & $0.968\pm 0.002$  &$31.1\pm 0.1$\\
      &Uninformed VDS~\cite{Lustig2007} & $0.925\pm 0.005$& $28.9\pm 0.1$\\
      \hline
       \textbf{Testing}&Our method  & $0.975\pm 0.003 $ & $32.1\pm 0.2$\\
      &Data-adapted~\cite{Knoll2011a} & $0.967\pm 0.003$ & $31.1\pm 0.2$\\
      &Uninformed VDS~\cite{Lustig2007} & $0.924\pm 0.003$& $28.8\pm 0.1$\\
      \hline
    \end{tabular}

  \end{table}
  \subsubsection{Cartesian line patterns}
  \label{sec:compare_line_patterns}
  Finally, we compare our method for Cartesian line patterns to another recent method for learning sampling patterns~\cite{Gozcu2018a} and uninformed variable density sampling patterns~\cite{Lustig2007}. In the method of~\cite{Gozcu2018a}, a set of candidate masks is considered and a sampling pattern is selected by adding candidate masks one at a time according to a greedy selection rule: at each stage, the candidate is chosen among the remaining candidates that gives the maximum increase of a performance measure on a training set. A drawback of the method from~\cite{Gozcu2018a} is that the lower level regularisation parameter, has to be fixed beforehand; we fix the regularisation parameter learned with our method on the training set, apply the method from~\cite{Gozcu2018a} to learn a line pattern, and finally tune the regularisation parameter on the training set with our method to improve the performance of the pattern learned with with the method from~\cite{Gozcu2018a}. The uninformed variable density sampling pattern from~\cite{Lustig2007} does not fix the reconstruction method, so we use our method to learn the optimal regularisation parameter on the training set for this sampling pattern. We use a training set of 7 slices and test on 70 slices different to the ones used in training.
  \begin{figure}[!htb]
    \centering
    \includegraphics{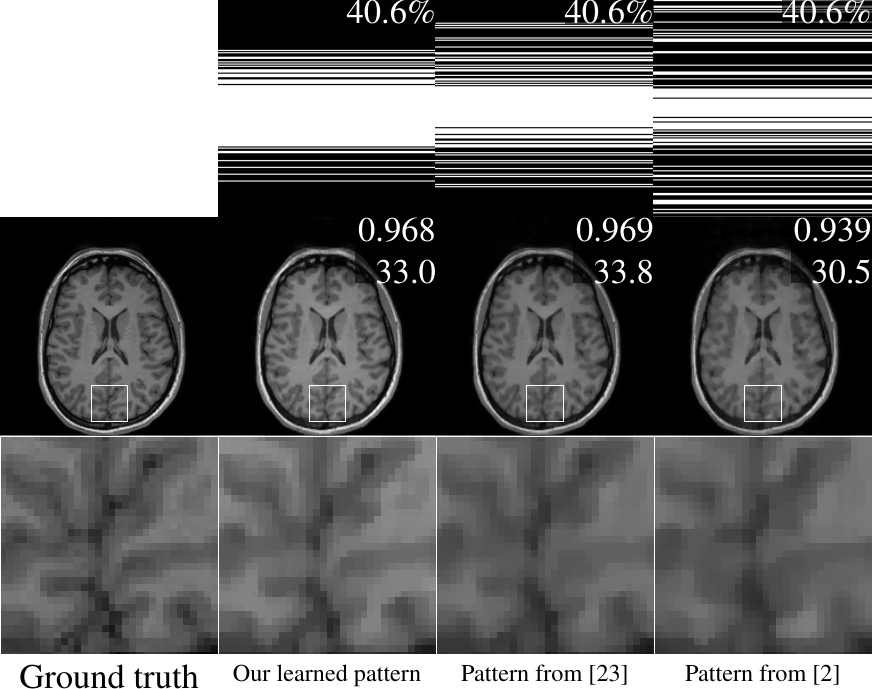}
    \caption{A comparison of our learned Cartesian line pattern to the learned pattern from~\cite{Gozcu2018a} and an uninformed variable density sampling pattern~\cite{Lustig2007} with TV regularisation in the lower level problem. On each of the reconstructions, the top number is the SSIM value and the bottom number is the PSNR.}
    \label{fig:line_comparison}
  \end{figure}

    \begin{table}[!h]
  \caption{A comparison of the performance of our learned Cartesian line pattern to the learned patterns of~\cite{Gozcu2018a} and uninformed variable density sampling patterns from~\cite{Lustig2007} with TV regularisation in the lower level problem. All compared sampling patterns sample 40.6\% of k-space.}
  \label{tab:compare_line_patterns}
  \centering

    \begin{tabular}{|l|l|l|l|}
      \hline
      &Pattern type& SSIM & PSNR\\
      \hline
      \textbf{Training}&Our method & $0.978\pm 0.002$ & $29.6\pm0.4$\\
      &Learned~\cite{Gozcu2018a}  & $0.980\pm 0.002$  &$30.5\pm 0.5$\\
      &Uninformed VDS~\cite{Lustig2007} & $0.959\pm0.005$& $28.2\pm 0.6$\\
      \hline
       \textbf{Testing}&Our method  & $0.969\pm 0.003 $ & $33.5\pm 0.9$\\
      &Learned~\cite{Gozcu2018a} & $0.969\pm 0.003$ & $34.2\pm 0.7$\\
      &Uninformed VDS~\cite{Lustig2007} & $0.944\pm 0.007$& $31.6\pm 0.7$\\
      \hline
    \end{tabular}

  \end{table}
  As we see in Figure~\ref{fig:line_comparison} and Table~\ref{tab:compare_line_patterns}, both our learned pattern and the learned pattern from~\cite{Gozcu2018a} significantly outperform the uninformed variable density sampling pattern from~\cite{Lustig2007}. Our learned pattern performs very similarly to the pattern from~\cite{Gozcu2018a}, if ever so slightly worse in terms of the performance metrics. A comparison of the computational effort required for the method in \cite{Gozcu2018a} and our method can be given by noting that the effort required in both methods is proportional to the number of times a lower level problem has to be solved. In our method, there is at each iteration an additional adjoint equation that needs to be solved, which takes less than but comparable effort to one lower level solve. That is, one iteration of our method effectively requires (less than) two lower level solves. For the method in \cite{Gozcu2018a}, assuming a set of $N$ candidate masks (disjoint and each of the same size) and a sampling rate $r$, we need to perform
\[
    \sum\limits_{i=0}^{rN}(N - i)=r\Big(1-\frac{r}{2}\Big)N^2 + \Big(1 -  \frac{r}{2}\Big)N = \Theta(N^2).
\]
    lower level solves. Table \ref{tab:computational_effort} shows two concrete settings in which we compare the computational effort (in terms of effective number of lower level solves) required to use each method.

    \begin{table}[!h]
      \centering
      \begin{tabular}{| l | l| l |}
        \hline
        & Line sampling (40.6\%) & Free pattern (34.7\%) \\\hline
        Our method & 4192 & 6494\\
        The method from \cite{Gozcu2018a} & 12087 & $3.90\cdot 10^8$  \\\hline
        
      \end{tabular}
      \caption{A comparison of the computational efforts (measured in effective number of lower level solves) required for our method and for the method in~\cite{Gozcu2018a} on images of size $192\times 192$. }
      \label{tab:computational_effort}
    \end{table}

    Note that we did not actually use the method in~\cite{Gozcu2018a} to learn a free pattern, since the number of lower level solves required to do this was prohibitive. By using a continuous optimisation approach to learning sampling patterns, our method can be more easily scaled up to higher resolutions and more computationally demanding settings such as 3D MRI or dynamic MRI; Quasi-Newton methods, such as the L-BFGS-B algorithm, exhibit a resolution independent behavior for problems like Problem~\eqref{prob:sampling_bilevel_prob} i.e., the number of outer iterations remains almost the same no matter the size of the variables involved~\cite{Kelley1987}.
    \subsection{High resolution example}
\label{sec:hires_example}
\begin{figure}[!htb]\centering
  \includegraphics[scale=1]{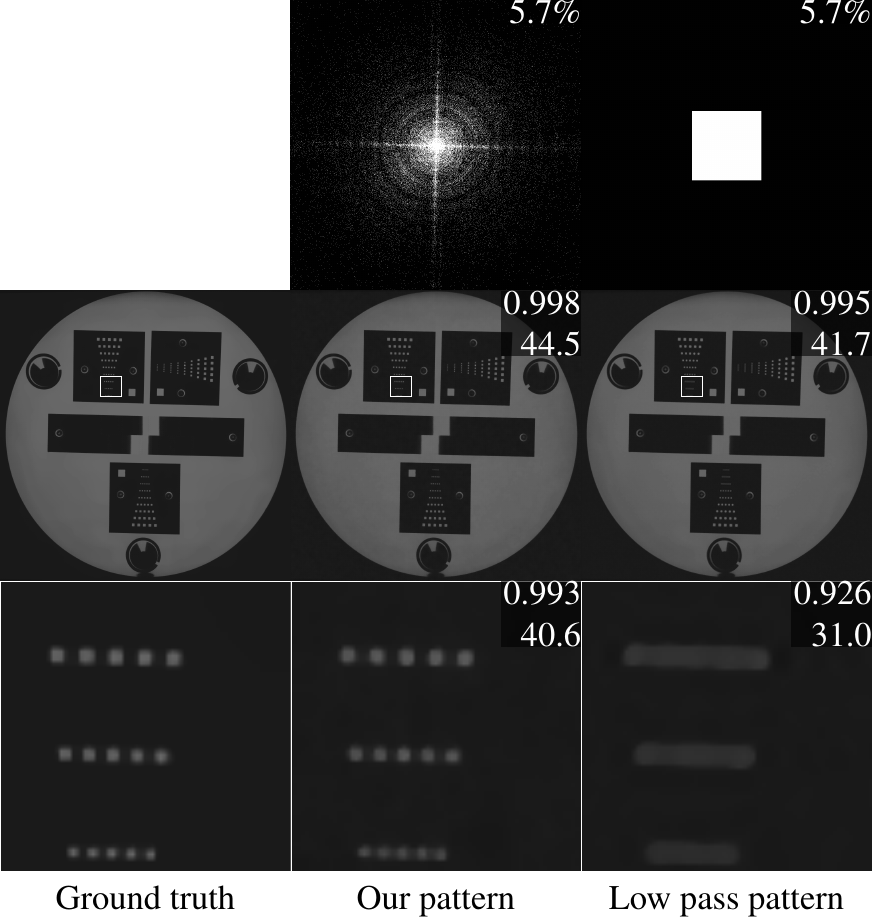}
  \caption{A comparison of the learned pattern and a low-pass sampling pattern in the high resolution setting with TV regularisation in the lower level problem. On each of the reconstructions, the top number is the SSIM value and the bottom number is the PSNR. On the bottom row, the performance metrics are computed using just the zoomed regions.}
  \label{fig:hires}
  \end{figure}

Up to this point, the experiments have been run on relatively small images. For this experiment, we used a training set of 5 slices taken from a high resolution scan of a phantom. In Figure~\ref{fig:hires}, we consider a different test slice from this scan to see how well the learned pattern performs. We compare our learned pattern to a low-pass sampling pattern (with the lower level regularisation parameter learned on the training set). Though both methods do well at reconstructing the phantom image, the zoomed region shows that our method allows fine details to be resolved very well, even when sampling just 5.7\% of k-space, whereas the low pass pattern has a limited resolution.
c


\section{Discussion and Outlook}
All our experiments were carried out on 2D images. With minor mathematical modifications, the proposed method can be applied to learn sampling patterns for 3D MRI, though it is worth noting that the computational effort will scale up accordingly and the implementation will need to be optimised to deal with this. There is considerable scope for optimisation of the computational implementation of the method, for instance by parallelising the solution method for the lower level problems. To accelerate MRI in practice, it is necessary to take into account the physical constraints imposed on sampling. The free pattern of points learned by our method is not immediately useful for accelerating 2D MRI, but it can be used for accelerated 3D MRI. If our method is extended to 3D MRI, the problem of efficiently sampling along these patterns in practice comes up again. In \cite{Boyer2016}, a method is proposed (which has been implemented in practice at NeuroSpin \cite{Lazarus2019}) that can be used to generate practical sampling strategies from a given target density. We can estimate a target density from our learned pattern, and use it as an input to this method.

Besides these extensions to our method, one can consider more general lower level regularisation functionals and allow for more flexibility to learn a custom regulariser as was done for denoising in \cite{Chen2014}, or unroll the lower level algorithm and use an approach similar to that of the variational network \cite{Hammernik2018}.

In this work, we have considered the free and Cartesian line parametrisations of the sampling pattern, but we mentioned that any differentiable parametrisation of the sampling pattern can be used. With an appropriate choice of the parametrisation, our method can be used to learn optimal radial line patterns, or other physically feasible optimal sampling patterns.

In our framework, we made smoothness assumptions on the lower level problems in order to differentiate their solution maps. Similar results can be derived assuming partial smoothness of the regularisation functionals \cite{Vaiter2017}, which covers total variation regularisation and the wavelet regularisation without needing to smooth them. The non-smooth lower level problems will be harder to solve, but it might be possible to deal directly with non-smooth lower level problems using this approach. Alternatively, one could consider optimality conditions for bilevel optimisation problems with non-smooth lower level problems \cite{Dempe2011} and attempt to solve the optimality conditions.

Despite being a smooth optimisation problem, the learning procedure is computationally intensive, since the lower level problems have to be solved to high accuracy in each iteration. These issues are alleviated by warm-starting the lower level solvers and it may be possible to do something similar with the iterative solver used to compute gradients. There is considerable scope for investigating ways in which the optimisation can be improved: the problem is non-convex so one could further research whether this is problematic in this case, and, if so, how to get around these issues. In Section~\ref{sec:cartesian_line}, we saw that, even with the penalty in the upper level that encourages discreteness of the learned patterns, the learned Cartesian line patterns were not binary, which may be an artefact of the difficulties involved in solving the optimisation problem. One thing that can be of great importance in non-convex optimisation is the initialisation that is used; in this work we have used a fixed initialisation consisting of an identity sampling operator and the corresponding optimal regularisation parameter and found that it generally worked well, but more detailed study may point to a more suitable initialisation. Since the objective function splits as a sum over the training set, another natural direction of future research would be to investigate the use of stochastic optimisation methods in this setting.
\section{Conclusions}
We have proposed a supervised learning approach to learn high quality sampling patterns for accelerated MRI for a given variational reconstruction method. We have demonstrated that this approach is highly flexible, allowing for a wide variety of regularisation functionals to be used and allowing constraints to be imposed on the learned sampling patterns.
Furthermore, we have shown that the method can be used successfully with small training sets. The learned patterns perform favourably compared to standard choices of sampling patterns, both quantitatively (measured by SSIM and PSNR on a test set) and qualitatively (by comparing the resolution of fine scale details).

This work shows that it is feasible to learn sampling patterns by applying continuous optimisation methods to a bilevel optimisation problem and suggests multiple ways in which this methodology can be extended to work in different settings.


\appendix
\label{app:details_app}
\subsection{Alternative parametrisations of the sampling pattern}
\label{sec:alt_params}
As was mentioned before, it is possible to use various parametrisations of the sampling pattern. We implement this by allowing $p$ to depend smoothly on another parameter $\lambda$, through $p: B\to C$. This generalised parametrisation includes the following ones, which are used in the results of the main text:
  \begin{itemize}
    \item If we let $B=[0,1]^{n_1}\times[0,\infty)$ or $B = [0,1]^{n_2}\times [0,\infty)$, and we let $p$ encode horizontal or vertical lines in k-space using the first $n_1$ or $n_2$ coordinates of $\lambda$ and the regularisation parameter with the last coordinate of $\lambda$, we can learn Cartesian line patterns and the regularisation parameter,
    \item If we have a fixed sampling pattern $\samp = \diag(s_1,\ldots, s_{n_1\cdot n_2})$ and let $p(\lambda) = (s_1,\ldots, s_{n_1\cdot n_2}, \lambda)$ with $B = [0,\infty)$, we can learn the optimal regularisation parameter for the fixed pattern $\samp$.
  \end{itemize}
  Instead of studying a problem like Problem~\eqref{prob:sampling_bilevel_prob}, our problem now becomes
  \[\min_{\lambda\in B} \frac{1}{N}\sum\limits_{i=1}^N L_{u_i^*}(\hat u_i(p(\lambda)) + P(p(\lambda)).\]
  The same methodology that is used in the main text can be used to tackle this problem and we can use the chain rule to get the gradients that we need: $\lambda \mapsto P(p(\lambda))$ has gradient given by $\nabla P(p(\lambda)) Dp(\lambda)$, and using Equation~\eqref{eq:sol_map_grad}, we see that $\lambda \mapsto L_{u_i^*}(\hat u_i(p(\lambda)))$ has gradient
  \begin{align*}
    -Dp(\lambda)^*& D_{p, u} E_{y_i}(\hat u_i(p(\lambda)); p(\lambda))\\&([D_u^2 E_{y_i}(\hat u_i(p(\lambda)); p(\lambda))]^{-1}\nabla L_{u_i^*}(\hat u_i(p(\lambda)))^*)
    \end{align*}
\subsection{Gradient and Hessian of the lower level regularisation}\label{sec:grad_hessian_regulariser}
The regularisers that we consider in the lower level problems are twice continuously differentiable, and we can give explicit formulas for their gradients and for the action of their Hessians. Although we have a complex image forward model, when we speak of differentiability we mean differentiability with respect to the real and imaginary parts separately. Similarly, pixelwise products of complex quantities should here be interpreted as separate multiplication of the real and imaginary parts. We only need to compute the gradient and Hessian of $J(z) = J((z^1,\ldots, z^M)=\sum_i \rho(|(z^1, \ldots, z^M|)$. Indeed, the regulariser $R$ satisfies $R(u) = J(\mathcal A u) = J(\mathcal A u)$, so $DR(u) = \mathcal A^* DJ(\mathcal Au)$ and $D^2R(u) = \mathcal A^* D^2 J(\mathcal Au)\mathcal A$. We denote the real and imaginary parts of $z^j$ by $z_\text{real}^j$ and $z_\text{imag}^j$ respectively. Differentiating the sum that defines $J$ with respect to $z^j_{\text{real}, i}, z^j_{\text{imag}, i}$, we find that \begin{equation}
\frac{\partial J}{\partial z^{j}_{\text{comp},i}}(z) = \frac{\rho'(|z|_i)}{|z|_i}z^j_{\text{comp},i},\qquad \text{for comp}\in \{\text{real, imag}\}.\label{expr:regulariser_partial_deriv}
\end{equation}
We make notation less cumbersome by defining
\(\phi(x) = \rho'(x) / x\). Using Expression~\eqref{expr:regulariser_partial_deriv}, we see that 
\begin{equation}
DJ(z) = \phi(|z|)\cdot z.\label{regulariser_full_deriv}
\end{equation}
To get the Hessian of $J$, consider a component $(DJ(z))^p_{\text{comp}, i}$ and differentiate with respect to $z^q_{\text{comp'}, j}$:
\begin{align}
  \label{eq:regulariser_second_partial_deriv}
  \frac{\partial^2 J}{\partial z^q_{\text{comp'}, j} \partial z^p_{\text{comp}, i}}(z)&=\frac{\phi'(|z|_i)}{|z|_i}\delta_{i, j} z^q_{\text{comp'}, j}z^p_{\text{comp}, i}\nonumber\\
                                                                                      & \qquad+ \phi(|z|_i) \delta_{(i, p, \text{comp}), (j, q, \text{comp'})}.
\end{align}
To ease notation, we define
\[\psi(x) = \begin{cases}0\quad&\text{if}\quad x=0\\
\frac{\phi'(x)}{x}\quad &\text{if}\quad x > 0.
\end{cases}.\]
The action of $D^2J(z)$ on a vector $w$ can now be computed using Equation~\eqref{eq:regulariser_second_partial_deriv}:
\begin{align}
  \label{eq:regulariser_hessian_action}
 D^2J(z) w &= \psi(|z|)\cdot z \cdot \Big(\sum_{\substack{p = 1,\ldots, M\\\text{comp}\in \{\text{real},\text{imag}\}}} z^p_\text{comp}\cdot w^p_\text{comp}\Big)\nonumber \\&\qquad+ \phi(|z|)\cdot w
\end{align}
\subsection{Details of solving the lower level problems}
\label{sec:details_ll_solve}
In Section~\ref{sec:method_ll} of the main text, we show that the lower level energy functional $E_y$ takes the saddle-point structure that is exploited in PDHG. In this section, we describe the computations that need to be made to choose the parameters correctly and apply the algorithm.
\subsubsection{Proximal operator of $F_2$}\label{proximal-operator-of-j}
Given how $F_2$ is defined, its proximal operator can be computed by applying
pixelwise the proximal operator of
\(\xi:x=(x^1,\ldots, x^M)\mapsto \alpha(p)\rho(\sqrt{|x^1|^2 + \ldots + |x^M|^2})\).
The optimality condition defining the proximal operator tells us that \(\prox_{\tau \xi}(x^1,\ldots, x^M)\) is the unique \(\hat x\)
satisfying \[(1 + \tau\alpha(p)\phi(|\hat x|))\hat x = x.\] That is, \(\hat x\)
is a scalar multiple of \(x\). Taking norms of both sides of this
equation, we get an equation
\[(1+\tau\alpha(p)\phi(C))C = |x|,\]
which is explicitly solvable for our choices of lower level regularisations, for $|\hat x|$ in terms of $|x|$. Denoting its solution by $C(|x|, \tau)$, we find that 
\(\prox_{\tau \xi}(x)=\hat x = C(|x|,\tau)x/|x|\), and hence $\prox_{\tau F_2}(z)_i = \prox_{\tau \xi}(z_i) = C(|z_i|, \tau) z_i / |z_i|$.
\subsubsection{Choosing the parameters and putting the algorithm together}
\label{sec:ll_details}
To apply PDHG, we need to be able to compute proximal operators for $F^*$ and $G$. Since Moreau's identity gives an explicit expression relating the proximal operator of $F$ and of $F^*$, it suffices to compute the proximal operator of $F$. Furthermore, since $F$ is separable, we have $\prox_{\tau F}(v_1, v_2) = (\prox_{\tau F_1}(v_1), \prox_{\tau F_2}(v_2))$. In the previous subsection, we showed that we can explicitly compute $\prox_{\tau F_2}$. Considering the optimality condition defining $\prox_{\tau F_1}$ we find that
\begin{equation}
  \prox_{\tau F_1}(v) = \F^{-1}(I+\tau \samp(p)^2)^{-1}(\F u + \tau \samp(p)^2 y).
  \label{eq:prox_F1}
\end{equation}
  Note that $I+\tau \samp(p)^2$ is a diagonal matrix so that its inverse can be computed by a simple coordinate-wise product between vectors. Since $G(u) =  \eps \|u\|^2/2$, we have $\prox_{\tau G}(u) = u / (\eps\tau + 1)$.
  
  To choose appropriate step sizes, we note that $F$ is strongly smooth, since $F_1$ is (its Hessian is $\F^{-1}\samp(p)^2 \F$, the norm of which is bounded above by $\|\samp(p)^2\| = \max_{i=1,\ldots, n} p_i^2$) and $F_2$ is as well (with constant bounded by $c(p)$ as shown in Section~\ref{sec:smoothness_constant_pdhg}). Hence the smoothness constant of $F$ is bounded by $\eta := \max\{\max_{i=1,\ldots, n} p_i^2, c(p)\}$. Furthermore, $G$ is strongly convex with constant $\eps$. Finally, we need an estimate on $\|\mathcal K\|$: since $\mathcal K = (I, \mathcal A)$, we have $\|\mathcal K\|\leq \sqrt{1+ \|\mathcal A\|^2}$. In the examples we consider, $\|\mathcal A\|$ is known or can be estimated from above: when $\mathcal A=W$ is an orthogonal wavelet transform we have $\|\mathcal A\| = 1$, while when $\mathcal A=\nabla$ we use a standard discretisation for which it is well known that $\|\mathcal A\|\leq \sqrt{8}$ \cite{Chambolle2004}. In any case, we have $\|\mathcal A\|\leq L$ for some known $L>0$. Choosing our parameters as 
  \begin{align*}\mu &= 2\sqrt{\frac{\eps}{(1+ L^2)\eta}},\quad
    \tau = \frac{\mu}{2 \eps}, \quad
    \sigma = \frac{\mu \eta}{2},\quad
    \theta = \frac{1}{1+\mu},
  \end{align*}
  makes PDHG converge linearly \cite{Chambolle2011}.
\subsubsection{Computing the smoothness constant of $F_2$ for solving the lower level problems}\label{sec:smoothness_constant_pdhg}
To compute step sizes for PDHG that give a linearly convergent algorithm, we require an estimate of the smoothness constant of $F_2$. Recall that $F_2$ can be written as $F_2(z) = \alpha(p) J(z)$. The smoothness constant of \(J\) can be estimated by an upper bound on the operator norm of the Hessian. Using the triangle inequality, Equation~\eqref{eq:regulariser_hessian_action} tells us that 
\begin{align}
 \|D^2J(z) w\| &\leq \sum_{\substack{p = 1,\ldots, M\\\text{comp}\in \{\text{real},\text{imag}\}}}\Big\|\psi(|z|)\cdot z \cdot \Big( z^p_\text{comp}\cdot w^p_\text{comp}\Big)\Big\| \nonumber\\&\qquad+ \|\phi(|z|)\cdot w\|.\label{eq:hess_action}
\end{align}
Let us consider a term with index $(p, \text{comp})$ in the first sum:
\[\Big(\psi(|z|)\cdot z \cdot(z^p_\text{comp} \cdot w^p_\text{comp})\Big)^q_{\text{comp'}, i} = \psi(|z|_i) z^q_{\text{comp'},i}  z^p_{\text{comp}, i} w^p_{\text{comp}, i}.\]
Since $|z^q_{\text{comp'}, i}  z^p_{\text{comp},i}| \leq \frac{1}{2} (|z^q_{\text{comp'},i}|^2 + |z^q_{\text{comp},i}|^2) \leq \frac{1}{2}|z|_i^2$, we find that
\[
  |\psi(|z|_i) \cdot z_i\cdot(z^p_{\text{comp},i}\cdot w^p_{\text{comp}, i})|\leq \frac{1}{2}\sup_{x \geq 0} (|\psi(x)|x^2)|w^p_{\text{comp}, i}|.
\]
Now $|w^p_{\text{comp, i}}| \leq |w|_i$ and $\psi(x) x= \phi'(x)$ , so we conclude that
\begin{equation}
  \label{eq:smoothness_first_term}
  \Big\|\psi(|z|)\cdot z \cdot \Big( z^p_\text{comp}\cdot w^p_\text{comp}\Big)\Big\| \leq \frac{\sqrt{2M}}{2} \sup_{x\geq 0} (|\phi'(x)| x) \|w\|
\end{equation}
For the final term in Inequality~\eqref{eq:hess_action}, we can simply use the bound
\begin{equation}
  \label{eq:smoothness_second_term}
  \|\phi(|z|)\cdot w\|\leq \sup_{x\geq 0} |\phi(x)| \|w\|.
\end{equation}
Combining the above inequalities, we find that
\begin{equation}
  \label{eq:smoothness_constant}
  \|D^2 J(z) \| \leq \sqrt{2} M^{\frac{3}{2}} \sup_{x\geq 0} (|\phi'(x)| x) + \sup_{x\geq 0} |\phi(x)|,
\end{equation}
so the functional $J$ is $L$-smooth with
\[L = \sqrt{2} M^{\frac{3}{2}} \sup_{x\geq 0}(|\phi'(x)| x) + \sup_{x\geq 0} |\phi(x)|\]
and $F_2 = \alpha(p) J$ has smoothness constant bounded by $c(p) = \alpha(p) L$
  \subsection{Computing the action of the Hessian of the lower level energy functional}
  \label{sec:hessian_details}
  In this section, we compute the action of the Hessian of the lower level energy functionals. To prevent the expressions from becoming overly cumbersome, let us split $E$ into parts:
    \begin{align*}
      \E{y}{u}{p} &= E_{\text{data}}(u;p) + E_{\text{reg}}(u;p) + E_{\eps-\text{convex}}(u; p),
    \end{align*}
    with
    \begin{align*}
      E_{\text{data}}(u; p) &= \frac{1}{2} \|\samp(p) (\F  u - y)\|^2,\\
      E_{\text{reg}}(u; p) &= \alpha(p) J(\mathcal Au),\\
      E_{\eps-\text{convex}}(u; p) &= \frac{\eps}{2}\|u\|^2.
    \end{align*}
    We can differentiate each of these components with respect to $u$ (using the results shown in Section~\ref{sec:grad_hessian_regulariser} to differentiate $E_{\text{reg}}$) to give
    \begin{align*}
      D_u E_{\text{data}}(u; p) &=  \F^{-1}\samp(p)^2(\F  u -y),\\
      D_u E_{\text{reg}}(u; p) &= \alpha(p)\mathcal A^*(\phi(|\mathcal Au|)\cdot \mathcal Au),\\
      D_u E_{\eps-\text{convex}}(u; p) &= \eps u.
    \end{align*}
Differentiating once again with respect to $u$ (again using the results in Section~\ref{sec:grad_hessian_regulariser}), we find that the actions of the various parts of the Hessian on a vector $w$ are given by
\begin{align*}
  D^2_u E_{\text{data}}(u; p)w& =  \F^{-1} \samp(p)^2 \F  w,\\
  D^2_u E_{\text{reg}}(u; p)&w= \alpha(p) \cdot \mathcal A^*\Big(\psi(|\mathcal A u|)\cdot \mathcal A u \cdot
  \\&\Big( \sum_{\substack{p=1,\ldots, M\\ \text{comp}\in \{\text{real, imag}\}}}(\mathcal Au)^p_\text{comp}\cdot (\mathcal Aw)^p_\text{comp}\Big)\\
                              &\qquad+\phi(|\mathcal A u|)\cdot \mathcal A w \Big),\\
  D^2_u E_{\eps-\text{convex}}(u; p) w &= \eps w.
\end{align*}
In addition to this, according to Equation~\eqref{eq:sol_map_grad}, we need access to $D_{p, u}$. Noting that $E_{\eps-\text{convex}}$ does not depend on $p$, we find that $D_{p, u}E_y$ acts on a vector $w$ as
\begin{align*}
  & (D_{p, u} \E{y}{u}{p}w)_{i} =\\
  &\quad\sum_{\text{comp}\in \{\text{real},\text{imag}\}}(\F w)_{\text{comp}, i} \cdot 2 p_i\cdot (\F u - y)_{\text{comp}, i},
  \end{align*}
 for $1\leq i\leq n$, and
 \[ (D_{p, u} \E{y}{u}{p}w)_{n + 1}  = w^* \mathcal A^*(\phi(|\mathcal A u|)\cdot \mathcal A u).\]


\bibliographystyle{ieeetran}
\bibliography{references}

\begin{thebibliography}{10}
\providecommand{\url}[1]{#1}
\csname url@samestyle\endcsname
\providecommand{\newblock}{\relax}
\providecommand{\bibinfo}[2]{#2}
\providecommand{\BIBentrySTDinterwordspacing}{\spaceskip=0pt\relax}
\providecommand{\BIBentryALTinterwordstretchfactor}{4}
\providecommand{\BIBentryALTinterwordspacing}{\spaceskip=\fontdimen2\font plus
\BIBentryALTinterwordstretchfactor\fontdimen3\font minus
  \fontdimen4\font\relax}
\providecommand{\BIBforeignlanguage}[2]{{%
\expandafter\ifx\csname l@#1\endcsname\relax
\typeout{** WARNING: IEEEtran.bst: No hyphenation pattern has been}%
\typeout{** loaded for the language `#1'. Using the pattern for}%
\typeout{** the default language instead.}%
\else
\language=\csname l@#1\endcsname
\fi
#2}}
\providecommand{\BIBdecl}{\relax}
\BIBdecl

\bibitem{Candes2006}
E.~J. Cand{\`{e}}s, J.~K. Romberg, and T.~Tao, ``{Stable signal recovery from
  incomplete and inaccurate measurements},'' \emph{Communications on Pure and
  Applied Mathematics}, vol.~59, no.~8, pp. 1207--1223, 2006.

\bibitem{Lustig2007}
M.~Lustig, D.~Donoho, and J.~M. Pauly, ``{Sparse MRI: The application of
  compressed sensing for rapid MR imaging},'' \emph{MRM}, vol.~58, no.~6, pp.
  1182--1195, 2007.

\bibitem{Sodickson2015}
D.~K. Sodickson \emph{et~al.}, ``{The rapid imaging renaissance: sparser
  samples, denser dimensions, and glimmerings of a grand unified tomography},''
  in \emph{Proceedings of SPIE}, B.~Gimi and R.~C. Molthen, Eds., vol.
  9417.\hskip 1em plus 0.5em minus 0.4em\relax International Society for Optics
  and Photonics, 2015, p. 94170G.

\bibitem{Ravishankar2011a}
S.~Ravishankar and Y.~Bresler, ``{MR Image Reconstruction From Highly
  Undersampled k-Space Data by Dictionary Learning},'' \emph{IEEE TMI},
  vol.~30, no.~5, pp. 1028--1041, 2011.

\bibitem{Ehrhardt2016}
M.~J. Ehrhardt and M.~M. Betcke, ``{Multicontrast MRI Reconstruction with
  Structure-Guided Total Variation},'' \emph{SIIMS}, vol.~9, no.~3, pp.
  1084--1106, 2016.

\bibitem{Lingala2011}
S.~G. Lingala \emph{et~al.}, ``{Accelerated dynamic MRI exploiting sparsity and
  low-rank structure: k-t SLR},'' \emph{IEEE TMI}, vol.~30, no.~5, pp.
  1042--1054, 2011.

\bibitem{Tremoulheac2014}
B.~Tr{\'{e}}moulh{\'{e}}ac \emph{et~al.}, ``{Dynamic MR Image
  Reconstruction-Separation From Undersampled (k, t)-Space via Low-Rank Plus
  Sparse Prior},'' \emph{IEEE TMI}, vol.~33, no.~8, pp. 1689--1701, 2014.

\bibitem{Piccini2011}
D.~Piccini \emph{et~al.}, ``{Spiral phyllotaxis: The natural way to construct a
  3D radial trajectory in MRI},'' \emph{MRM}, vol.~66, no.~4, pp. 1049--1056,
  2011.

\bibitem{Feng2014}
L.~Feng \emph{et~al.}, ``{Golden-angle radial sparse parallel MRI: Combination
  of compressed sensing, parallel imaging, and golden-angle radial sampling for
  fast and flexible dynamic volumetric MRI},'' \emph{MRM}, vol.~72, no.~3, pp.
  707--717, 2014.

\bibitem{Liu2014}
J.~Liu and D.~Saloner, ``{Accelerated MRI with CIRcular Cartesian UnderSampling
  (CIRCUS): a variable density Cartesian sampling strategy for compressed
  sensing and parallel imaging.}'' \emph{QIMS}, vol.~4, no.~1, pp. 57--67,
  2014.

\bibitem{Paquette2015}
M.~Paquette \emph{et~al.}, ``{Comparison of sampling strategies and sparsifying
  transforms to improve compressed sensing diffusion spectrum imaging},''
  \emph{MRM}, vol.~73, no.~1, pp. 401--416, 2015.

\bibitem{Usman2009}
M.~Usman and P.~G. Batchelor, ``{Optimized Sampling Patterns for Practical
  Compressed MRI},'' \emph{SampTA'09}, 2009.

\bibitem{Puy2011}
G.~Puy, P.~Vandergheynst, and Y.~Wiaux, ``{On Variable Density Compressive
  Sampling},'' \emph{IEEE Signal Processing Letters}, vol.~18, no.~10, pp.
  595--598, 2011.

\bibitem{Chauffert2013}
N.~Chauffert, P.~Ciuciu, and P.~Weiss, ``{Variable density compressed sensing
  in MRI. Theoretical vs heuristic sampling strategies},'' in \emph{2013 IEEE
  10th International Symposium on Biomedical Imaging}.\hskip 1em plus 0.5em
  minus 0.4em\relax IEEE, 2013, pp. 298--301.

\bibitem{Chauffert2014}
N.~Chauffert \emph{et~al.}, ``{Variable Density Sampling with Continuous
  Trajectories},'' \emph{SIAM Journal on Imaging Sciences}, vol.~7, no.~4, pp.
  1962--1992, 2014.

\bibitem{Adcock2017}
B.~Adcock \emph{et~al.}, ``{Breaking the coherence barrier: a new theory for
  compressed sensing},'' \emph{Forum of Mathematics, Sigma}, vol.~5, p.~e4,
  2017.

\bibitem{Krahmer2014}
F.~Krahmer and R.~Ward, ``{Stable and Robust Sampling Strategies for
  Compressive Imaging},'' \emph{IEEE TIP}, vol.~23, no.~2, pp. 612--622, 2014.

\bibitem{Poon2016}
C.~Poon, ``{On Cartesian line sampling with anisotropic total variation
  regularization},'' 2016.

\bibitem{Boyer2019}
C.~Boyer, J.~Bigot, and P.~Weiss, ``{Compressed sensing with structured
  sparsity and structured acquisition},'' \emph{Applied and Computational
  Harmonic Analysis}, vol.~46, no.~2, pp. 312--350, 2019.

\bibitem{Ravishankar2011}
S.~Ravishankar and Y.~Bresler, ``{Adaptive sampling design for compressed
  sensing MRI},'' in \emph{2011 Annual International Conference of the IEEE
  EMBS}.\hskip 1em plus 0.5em minus 0.4em\relax IEEE, 2011, pp. 3751--3755.

\bibitem{Seeger2009}
M.~Seeger \emph{et~al.}, ``{Optimization of k-space trajectories for compressed
  sensing by Bayesian experimental design},'' \emph{MRM}, vol.~63, no.~1, pp.
  116--126, 2009.

\bibitem{Knoll2011}
F.~Knoll \emph{et~al.}, ``{Second order total generalized variation (TGV) for
  MRI},'' \emph{MRM}, vol.~65, no.~2, pp. 480--491, 2011.

\bibitem{Gozcu2018a}
B.~G\"ozc\"u \emph{et~al.}, ``{Learning-Based Compressive MRI},'' \emph{IEEE
  TMI}, vol.~37, no.~6, pp. 1394--1406, 2018.

\bibitem{Gozcu2019}
B.~G{\"{o}}zc{\"{u}}, T.~Sanchez, and V.~Cevher, ``{Rethinking Sampling in
  Parallel MRI: A Data-Driven Approach},'' in \emph{27th European Signal
  Processing Conference (EUSIPCO)}, 2019, pp. 1--5.

\bibitem{Haldar2019}
J.~P. Haldar and D.~Kim, ``{OEDIPUS: An Experiment Design Framework for
  Sparsity-Constrained MRI},'' \emph{IEEE TMI}, pp. 1--1, 2019.

\bibitem{Weiss2019}
T.~Weiss \emph{et~al.}, ``{Learning Fast Magnetic Resonance Imaging},'' 2019.

\bibitem{Li2016}
Y.-H. Li and V.~Cevher, ``{Learning data triage: Linear decoding works for
  compressive MRI},'' in \emph{2016 IEEE ICASSP}.\hskip 1em plus 0.5em minus
  0.4em\relax IEEE, 2016, pp. 4034--4038.

\bibitem{Delosreyes2017}
J.~C. {De los Reyes}, C.-B. Sch{\"{o}}nlieb, and T.~Valkonen, ``{Bilevel
  Parameter Learning for Higher-Order Total Variation Regularisation Models},''
  \emph{JMIV}, vol.~57, no.~1, pp. 1--25, 2017.

\bibitem{Huber1964}
P.~J. Huber, ``{Robust Estimation of a Location Parameter},'' \emph{The Annals
  of Mathematical Statistics}, vol.~35, no.~1, pp. 73--101, 1964.

\bibitem{Rudin1992}
L.~I. Rudin, S.~Osher, and E.~Fatemi, ``{Nonlinear total variation based noise
  removal algorithms},'' \emph{Physica D}, vol.~60, pp. 259--268, 1992.

\bibitem{Guerquin-Kern2009}
M.~Guerquin-Kern \emph{et~al.}, ``{Wavelet-regularized reconstruction for rapid
  MRI},'' in \emph{2009 IEEE International Symposium on Biomedical Imaging:
  From Nano to Macro}.\hskip 1em plus 0.5em minus 0.4em\relax IEEE, 2009, pp.
  193--196.

\bibitem{Pejoski2015}
S.~Pejoski, V.~Kafedziski, and D.~Gleich, ``{Compressed Sensing MRI Using
  Discrete Nonseparable Shearlet Transform and FISTA},'' \emph{IEEE Signal
  Processing Letters}, vol.~22, no.~10, pp. 1566--1570, 2015.

\bibitem{Chambolle2011}
A.~Chambolle and T.~Pock, ``{A First-Order Primal-Dual Algorithm for Convex
  Problems with Applications to Imaging},'' \emph{JMIV}, vol.~40, no.~1, pp.
  120--145, 2011.

\bibitem{Byrd1995a}
R.~H. Byrd \emph{et~al.}, ``{A Limited Memory Algorithm for Bound Constrained
  Optimization},'' \emph{SISC}, vol.~16, no.~5, pp. 1190--1208, 1995.

\bibitem{Zhu1997}
C.~Zhu \emph{et~al.}, ``{Algorithm 778: L-BFGS-B: Fortran subroutines for
  large-scale bound-constrained optimization},'' \emph{ACM Transactions on
  Mathematical Software}, vol.~23, no.~4, pp. 550--560, 1997.

\bibitem{Paszke2019}
A.~Paszke \emph{et~al.}, ``{PyTorch: An Imperative Style, High-Performance Deep
  Learning Library},'' in \emph{Advances in Neural Information Processing
  Systems 32 (NIPS 2019)}, 2019, pp. 8026--8037.

\bibitem{2020SciPy-NMeth}
P.~Virtanen \emph{et~al.}, ``{SciPy 1.0: Fundamental Algorithms for Scientific
  Computing in Python},'' \emph{Nature Methods}, vol.~17, pp. 261--272, 2020.

\bibitem{fergal_cotter_fbcotterpytorch_wavelets_2019}
F.~Cotter and S.~McLaughlin, ``fbcotter/pytorch\_wavelets: {Zenodo}
  {Release},'' Oct. 2019.

\bibitem{Cocosco1997}
C.~A. Cocosco \emph{et~al.}, ``{BrainWeb: Online Interface to a 3D MRI
  Simulated Brain Database},'' \emph{NEUROIMAGE}, vol.~5, p. 425, 1997.

\bibitem{Benning2014}
M.~Benning \emph{et~al.}, ``{Phase reconstruction from velocity-encoded MRI
  measurements - A survey of sparsity-promoting variational approaches},''
  \emph{JMR}, vol. 238, pp. 26--43, 2014.

\bibitem{Knoll2011a}
F.~Knoll \emph{et~al.}, ``{Adapted random sampling patterns for accelerated
  MRI},'' \emph{MAGMA}, vol.~24, no.~1, pp. 43--50, feb 2011.

\bibitem{Kelley1987}
C.~T. Kelley and E.~W. Sachs, ``{Quasi-Newton Methods and Unconstrained Optimal
  Control Problems},'' \emph{SIAM Journal on Control and Optimization},
  vol.~25, no.~6, pp. 1503--1516, nov 1987.

\bibitem{Boyer2016}
C.~Boyer \emph{et~al.}, ``{On the Generation of Sampling Schemes for Magnetic
  Resonance Imaging},'' \emph{SIIMS}, vol.~9, no.~4, pp. 2039--2072, 2016.

\bibitem{Lazarus2019}
C.~Lazarus \emph{et~al.}, ``{SPARKLING: variable-density k-space filling curves
  for accelerated $T_2^*$-weighted MRI},'' \emph{MRM}, vol.~81, no.~6, pp.
  3643--3661, 2019.

\bibitem{Chen2014}
Y.~Chen, ``{Learning fast and effective image restoration models},'' Ph.D.
  dissertation, Graz University of Technology, 2014.

\bibitem{Hammernik2018}
K.~Hammernik \emph{et~al.}, ``{Learning a variational network for
  reconstruction of accelerated MRI data},'' \emph{MRM}, vol.~79, no.~6, pp.
  3055--3071, 2018.

\bibitem{Vaiter2017}
S.~Vaiter \emph{et~al.}, ``{The degrees of freedom of partly smooth
  regularizers},'' \emph{AISM}, vol.~69, no.~4, pp. 791--832, 2017.

\bibitem{Dempe2011}
S.~Dempe and A.~B. Zemkoho, ``{The Generalized Mangasarian-Fromowitz Constraint
  Qualification and Optimality Conditions for Bilevel Programs},'' \emph{JOTA},
  vol. 148, no.~1, pp. 46--68, 2011.

\bibitem{Chambolle2004}
A.~Chambolle, ``{An Algorithm for Total Variation Minimization and
  Applications},'' \emph{JMIV}, vol.~20, pp. 89--97, 2004.

\end{thebibliography}

\end{document}


\algblock{ParFor}{EndParFor}

\algnewcommand\algorithmicparfor{\textbf{parfor}}
\algnewcommand\algorithmicpardo{\textbf{do}}
\algnewcommand\algorithmicendparfor{\textbf{end\ parfor}}
\algrenewtext{ParFor}[1]{\algorithmicparfor\ #1\ \algorithmicpardo}
\algrenewtext{EndParFor}{\algorithmicendparfor}
\title{Supporting document for "Learning the Sampling Pattern for MRI"}
\author{Ferdia~Sherry, Martin~Benning, Juan~Carlos~De~los~Reyes, Martin~J.~Graves, Georg~Maierhofer, Guy~Williams, Carola-Bibiane~Sch\"onlieb and Matthias~J.~Ehrhardt}

\maketitle\old{
\section{Computing the smoothness constant of $F_2$ for solving the lower level problems}\label{sec:smoothness_constant_pdhg}
To compute step sizes for PDHG that give a linearly convergent algorithm, we require an estimate of the smoothness constant of $F_2$. Recall that $F_2$ can be written as
\[F_2\begin{pmatrix} z^1\\\vdots \\z^M\end{pmatrix} = \alpha(p)\sum\limits_i \rho\Big(\sqrt{(z^1_i)^2 + \ldots + (z^M_i)^2}\Big).\]
This functional is a constant multiple of a functional of the form $R$ studied in Section~B of the Appendix of the main text if we take
\begin{align*}
\mathcal A_1\begin{pmatrix}z^1\\\vdots\\z^N\end{pmatrix} = \begin{pmatrix} z^1\\0\\\vdots\\0\end{pmatrix}, \ldots, \mathcal A_M \begin{pmatrix} z^1\\\vdots\\z^M\end{pmatrix} = \begin{pmatrix}z^M\\0\\\vdots\\0\end{pmatrix}.
\end{align*}
The smoothness constant of \(R\) can be estimated by an upper bound on the operator norm of the Hessian. Equation~(12) in the appendix of the main text tells us that
the Hessian is given by
\begin{align*}
D^2R(u) = \sum\limits_{p, q = 1}^M &\mathcal A_p^* \cdot(\psi(|\mathcal Au|)\cdot \mathcal A_p u\cdot \mathcal A_qu) \cdot \mathcal A_q \\&+ \sum\limits_{p=1}^M \mathcal A_p^*\cdot\phi(|\mathcal Au|)\cdot \mathcal A_p
\end{align*} For each \(i\) we have \(\|\mathcal A_i\| = \|\mathcal A_i^*\| = 1\) and the norms of
the multiplication operators in the above expression can be bounded as
follows: we have \begin{align*}
                   \Big|\psi\Big(&\sqrt{x_1^2 + \ldots + x_N^2}\Big) x_i x_j\Big|\\
                   &= \Big|\phi'\Big(\sqrt{x_1^2 + \ldots + x_N^2}\Big)\Big| \Big|\frac{x_i x_j}{\sqrt{x_1^2 + \ldots + x_N^2}}\Big|\\
&\leq \frac{1}{2}\Big|\phi'\Big(\sqrt{x_1^2 + \ldots + x_N^2}\Big)\Big|\sqrt{x_1^2 + \ldots + x_N^2}\\
&\leq \frac{1}{2}\sup_{x \geq 0} |\phi'(x)| x
\end{align*} so the norms of the inner parts of the sum defining
\(D^2R(u)\) (as linear operators, acting by pixelwise multiplication)
can be estimated as
\[\|\psi(|\mathcal Au|) \cdot \mathcal A_p u \cdot \mathcal A_q u\|\leq\frac{1}{2}\sup_{x\geq 0} |\phi'(x)| x\]
and \[\|\phi(|\mathcal Au|)\|\leq \sup_{x\geq 0} |\phi(x)|.\] Combining these estimates with the expression that we have for \(D^2R(u)\), we find that
\[\|D^2 R(u) \|\leq \frac{M^2}{2}\sup_{x\geq 0} (|\phi'(x)| x) + M \sup_{x\geq 0} |\phi(x)|,\]
so the functional \(R\) is \(L\)-smooth with \[L = \frac{M^2}{2}\sup_{x\geq 0}(|\phi'(x)| x) + M\sup_{x\geq 0} |\phi(x)|\]
and $F_2 = \alpha(p) R$ has smoothness constant bounded by $c(p) = \alpha(p)L$.}

\section{The test set}
\begin{figure}[!htb]
  \centering
  \includegraphics[width=\linewidth]{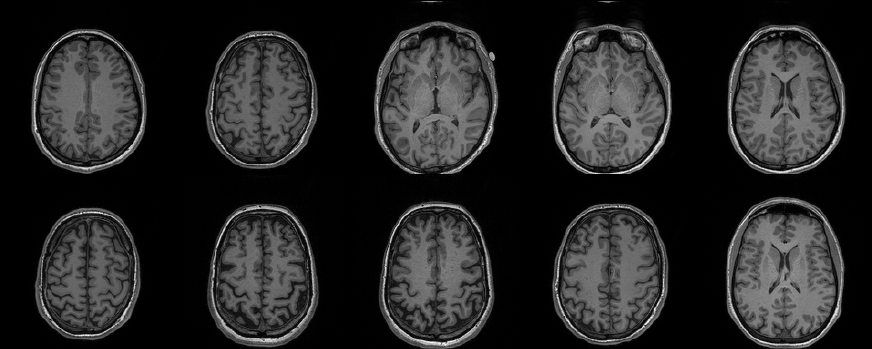}
  \caption{Examples of slices from the test set used to check the performance of the learned patterns.}
  \label{fig:test_set}
\end{figure}

Figure~\ref{fig:test_set} shows some of the variety in the images that were used to test the learned patterns.

\section{Varying the sparsity parameter $\beta$}
\begin{figure}[!htb]
  \centering
  \includegraphics[scale=1]{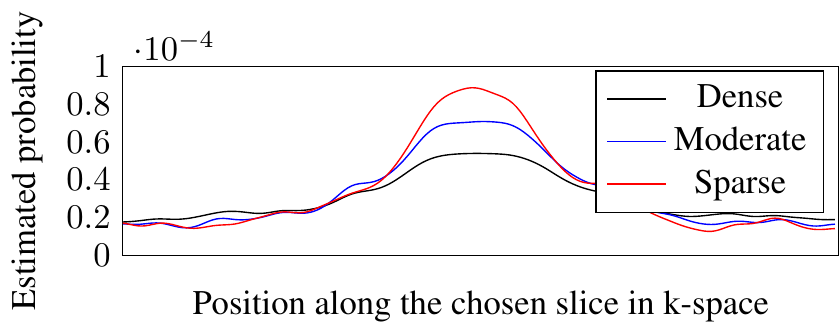}
  \caption{Slices through the centre of k-space for the Gaussian kernel density estimates of the sampling distributions for reconstruction with TV regularisation.}
  \label{fig:tv_plot_kde}
\end{figure}

Figure~\ref{fig:tv_plot_kde} shows slices through Gaussian kernel density estimates of the sampling distributions in Figure~4 in the main text.

\section{Varying the size of the training set}
\begin{figure}[!htb]
  \centering
  \includegraphics[scale=1]{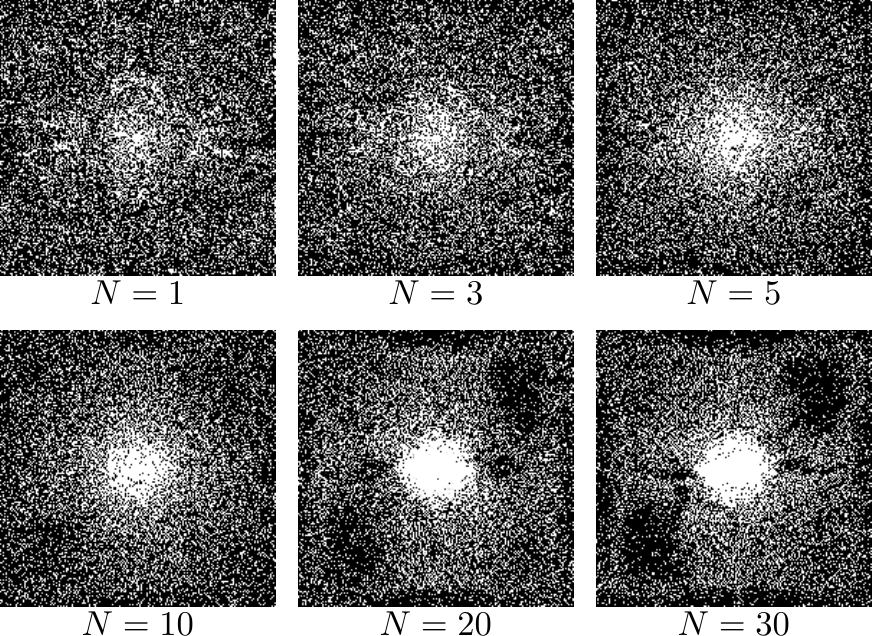}
  \caption{The learned sampling patterns as the size of the training set varies with TV regularisation in the lower level problem.}
  \label{fig:vary_training_patterns}
\end{figure}

Figure~\ref{fig:vary_training_patterns} shows the learned patterns for each of the training set sizes investigated in Section~III-E of the main text. It can be seen that the patterns become more strongly peaked around the centre of k-space as the size of the training set increases.

\section{Training set for high resolution example}
\label{sec:hires_training}
\begin{figure}[!htb]
    \centering
    \includegraphics[scale=1]{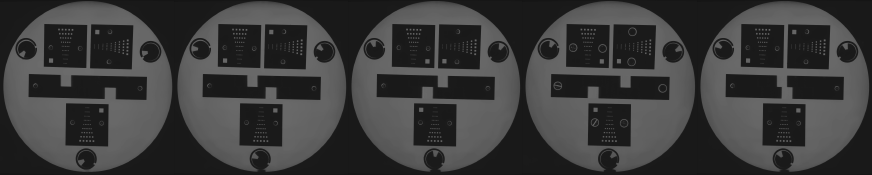}
    \caption{The slices that were used to train the high resolution sampling pattern.}
    \label{fig:hires_training}
  \end{figure}

  In the high resolution example of Section \ref{sec:hires_example} of the main text, we used the slices shown in Figure \ref{fig:hires_training} for the training set.
  \section{Convergence of  L-BFGS-B for the upper level problem}
\begin{figure}[!ht]
    \centering
    \includegraphics[scale=1]{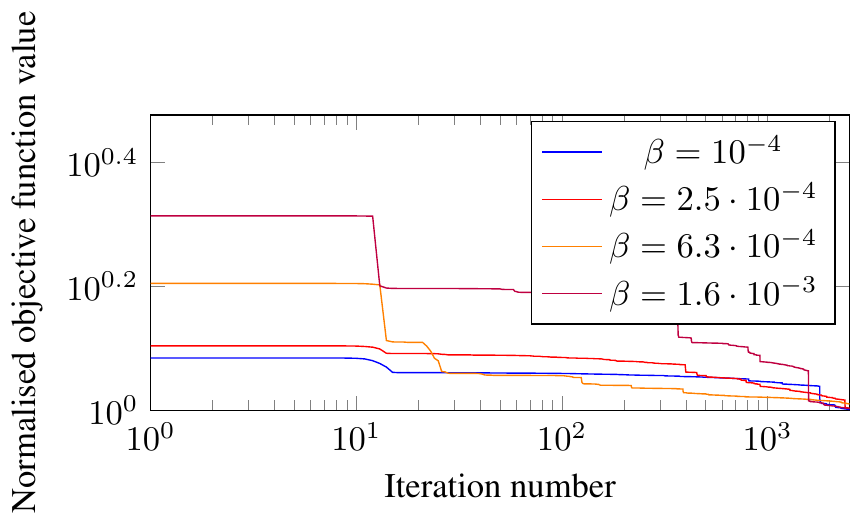}
    \caption{Plots showing the convergence of the upper level objective function value for a number of choices of $\beta$.}
    \label{fig:plot_lbfgsb}
  \end{figure}

  The plots in Figure \ref{fig:plot_lbfgsb} show the evolution of the (normalised) objective function value with the iteration number for some of the experiments in Section \ref{sec:vary_beta} of the main text.
